\documentclass[acmsmall,screen,nonacm]{acmart}

\usepackage[utf8]{inputenc}
\usepackage[T1]{fontenc}
\usepackage{microtype}
\usepackage[ruled, linesnumbered]{algorithm2e}
\usepackage{fancyhdr,graphicx,amsmath,eqnarray}
\usepackage{caption} 
\usepackage{balance}
\usepackage{subfigure}
\usepackage{multirow}
\usepackage{tikz}
\usepackage{pgfplots}
\usepackage{threeparttable}
\usepackage{booktabs}
\usepackage{enumitem}
\usepackage{makecell}
\usepackage[normalem]{ulem}
\usepackage{tabularx}
\usepackage{flushend}
\usepackage{xurl}
\usetikzlibrary{shapes, shapes.multipart, arrows.meta, positioning, matrix, calc, fit, decorations.pathreplacing, pgfplots.dateplot}
\usepgfplotslibrary{dateplot, statistics}
\pgfplotsset{compat=1.18}

\newcommand{\MIG}{{\texttt{Hagasu}}\xspace}

\begin{document}

\title{Chart the Invisible: Interactive Perception for Hidden-Target Exploration in GUI Agents}

\author{Yongxiang Hu}
\affiliation{%
  \institution{College of Computer Science and Artificial Intelligence, Fudan University}
  \city{Shanghai}
  \country{China}
}

\author{Ruiyang Liu}
\affiliation{%
  \institution{College of Computer Science and Artificial Intelligence, Fudan University}
  \city{Shanghai}
  \country{China}
}

\author{Qining He}
\affiliation{%
  \institution{College of Computer Science and Artificial Intelligence, Fudan University}
  \city{Shanghai}
  \country{China}
}

\author{Xuan Wang}
\affiliation{%
  \institution{College of Computer Science and Artificial Intelligence, Fudan University}
  \city{Shanghai}
  \country{China}
}

\author{Yangfan Zhou}
\affiliation{%
  \institution{College of Computer Science and Artificial Intelligence, Fudan University}
  \city{Shanghai}
  \country{China}
}

\begin{abstract}
Despite rapid progress in command-line interface (CLI) tasks, recent LLM-based agents still struggle with reliable task execution through graphical user interfaces (GUIs).
A key challenge is the limited content visible within a bounded viewport, especially on mobile devices, where many task-relevant targets remain hidden and can only be revealed through swipe interactions.
Therefore, although most GUI agents support swipe as an action primitive, the lack of target visibility makes it difficult for them to decide when, where, and how to swipe to reveal the hidden content.
Our quantitative study further shows that GUI agents incorrectly choose a click action in 47.8\% of cases that require swipe exploration.
To bridge this gap, this paper proposes equipping GUI agents with scrollable-region perception to support hidden-area exploration before action selection.
To this end, we trained the first multimodal model for scrollable-region perception on a manually collected dataset of 667 UI screens, called \MIG.
By correctly identifying 78.5\% of scrollable regions, experiments across nine GUI agents show that integrating \MIG improves the mean end-to-end completion rate from 64.8\% to 82.7\% on explore-dependent tasks.
To facilitate follow-up research, we release the swipe dataset and the source code of \MIG.
\end{abstract}



\maketitle

\section{Introduction}
\label{sec:intro}

Although large language models (LLMs) have made surprising progress on command-line interface (CLI) tasks~\cite{yang2023intercode,yang2024sweagent}, reliable task execution through graphical user interfaces (GUIs) remains challenging~\cite{xie2024osworld,kong2026mobileworld}. Beyond the well-known challenge of interpreting visual semantics, the limited content visible within a bounded viewport constitutes a key bottleneck for GUI agents. Specifically, to preserve visual appeal, UI designs constrain element density and thereby prevent visual clutter~\cite{reinecke2013aesthetics}. Modern UIs therefore widely use scrollable containers, such as lists and carousels, to accommodate additional functionality and content beyond the current view, requiring scrolling to reveal them~\cite{pfeuffer2018grid}. At every step, GUI agents are thus forced to decide under uncertainty whether to pick a visible target or explore hidden areas.

This dilemma is particularly acute in mobile GUIs.
Limited screen size leaves many UI elements outside the current viewport, accessible only through swipe interactions.
Although most GUI agents already support swipe as an action primitive~\cite{rawles2023aitw,zhang2023appagent,wang2024mobileagent,lin2024showui,qin2025uitars}, recent research shows that reliably completing mobile tasks remains challenging~\cite{rawles2025androidworld,kong2026mobileworld}.
To quantify this difficulty, this paper conducts a study of 7 GUI agents across 218 tasks and finds that the rate of correctly planned swipe actions is 47.3 percentage points lower than that of click actions.
Further analysis of 83.73\% failed tasks shows that a major cause of failure is clicking an incorrect visible target when the correct target is still hidden.
These findings highlight the need to help agents recognize opportunities for hidden-area exploration rather than force a choice among visible elements.

Therefore, this paper focuses on equipping GUI agents with scrollable-region perception, providing explicit cues about where hidden-area exploration is possible.
Achieving this goal involves two challenges.
First, scrollable regions lack clear visual indicators, requiring an understanding of UI structure and semantics to identify them.
To address this challenge, this paper builds on the UI understanding capabilities of a pretrained vision-language model (VLM) and fine-tunes it on 540 manually collected and annotated UI screens, to develop \MIG for scrollable region perception.
Second, improving a specific capability of a generalist GUI agent may compromise its broader capabilities.
This paper therefore implements \MIG as a non-intrusive plug-in that modifies only the GUI screenshot and prompt presented to the agent: it marks predicted scrollable regions on the screenshot and adds textual cues for swipe exploration.

Extensive experiments evaluate \MIG in scrollable-region perception, swipe planning, and end-to-end task completion.
Comparisons and ablations across eight configurations demonstrate the effectiveness of \MIG's training design for scrollable-region perception, with the full configuration achieving the highest recall of 78.46\%.
In offline swipe-planning experiments with nine VLMs, \MIG raises mean swipe prediction recall from 46.02\% to 53.87\%. And it decreases click prediction recall by only 1.67 percentage points.
Finally, online experiments show that integrating \MIG into existing GUI agents increases mean end-to-end task success rate from 64.81\% to 82.72\%.
To facilitate follow-up research, we release the scrollable-region dataset and the source code of \MIG.

We summarize the contributions of this paper as follows.
\begin{itemize}[leftmargin=*]
    \item We systematically identify off-screen content as a critical, understudied bottleneck that partially explains the performance gap between GUI and CLI agents.
    \item We build the first dataset of scrollable UI regions and open-source \MIG, a multimodal model that equips GUI agents with scrollable-region perception for proactive hidden-area exploration.
    \item Comprehensive experiments demonstrate the effectiveness of scrollable-region perception in improving GUI agents' task completion on explore-dependent tasks.
\end{itemize}

The rest of the paper is organized as follows.
Section~\ref{sec:background} presents the background and the preliminary study motivating this work.
Section~\ref{sec:method} describes the data collection and annotation, model fine-tuning, and plug-in design of \MIG.
Sections~\ref{sec:eva_design} and~\ref{sec:evaluation} present the experimental design and analyze the results, respectively.
Section~\ref{sec:threats} discusses threats to validity, and Section~\ref{sec:related} reviews related work.
Finally, Section~\ref{sec:conclusion} concludes the paper.

\section{Background}
\label{sec:background}

This section explains the decision pattern underlying ordinary click interactions and why hidden targets require a different course of action.
We first describe greedy interaction as selecting a task-relevant component from the current viewport.
We then explain why agents need swipe-based exploration when the correct target is hidden outside that viewport.
Finally, we present a preliminary study comparing click and swipe planning across different interaction scenarios.

\subsection{Greedy Interaction Patterns for Visible Targets}

GUI agents commonly follow an observation, decision, and execution loop, selecting actions based on the current interface, task instruction, and interaction history~\cite{yao2023react,zhang2023appagent,wang2024mobileagent,rawles2025androidworld}.
For a click action, this decision involves selecting an element in the current viewport and clicking it.
We describe this pattern as \emph{greedy interaction}: choosing the visible element that appears most likely to advance the task.
Here, greedy characterizes the choice among currently visible candidates, even when guided by an explicit plan or interaction history.

This strategy works when the current viewport contains the correct target, reducing click planning to semantic matching and coordinate grounding.
For example, an agent can fulfill a request to open Settings by identifying and clicking the visible Settings button.
Existing GUI grounding methods support this process by mapping task descriptions to visible UI elements and their locations~\cite{shaw2023pixels,baechler2024screenai,cheng2024seeclick,you2024ferretui,wu2025guiactor}.

\begin{figure}[t]
    \centering
    \includegraphics[width=\columnwidth]{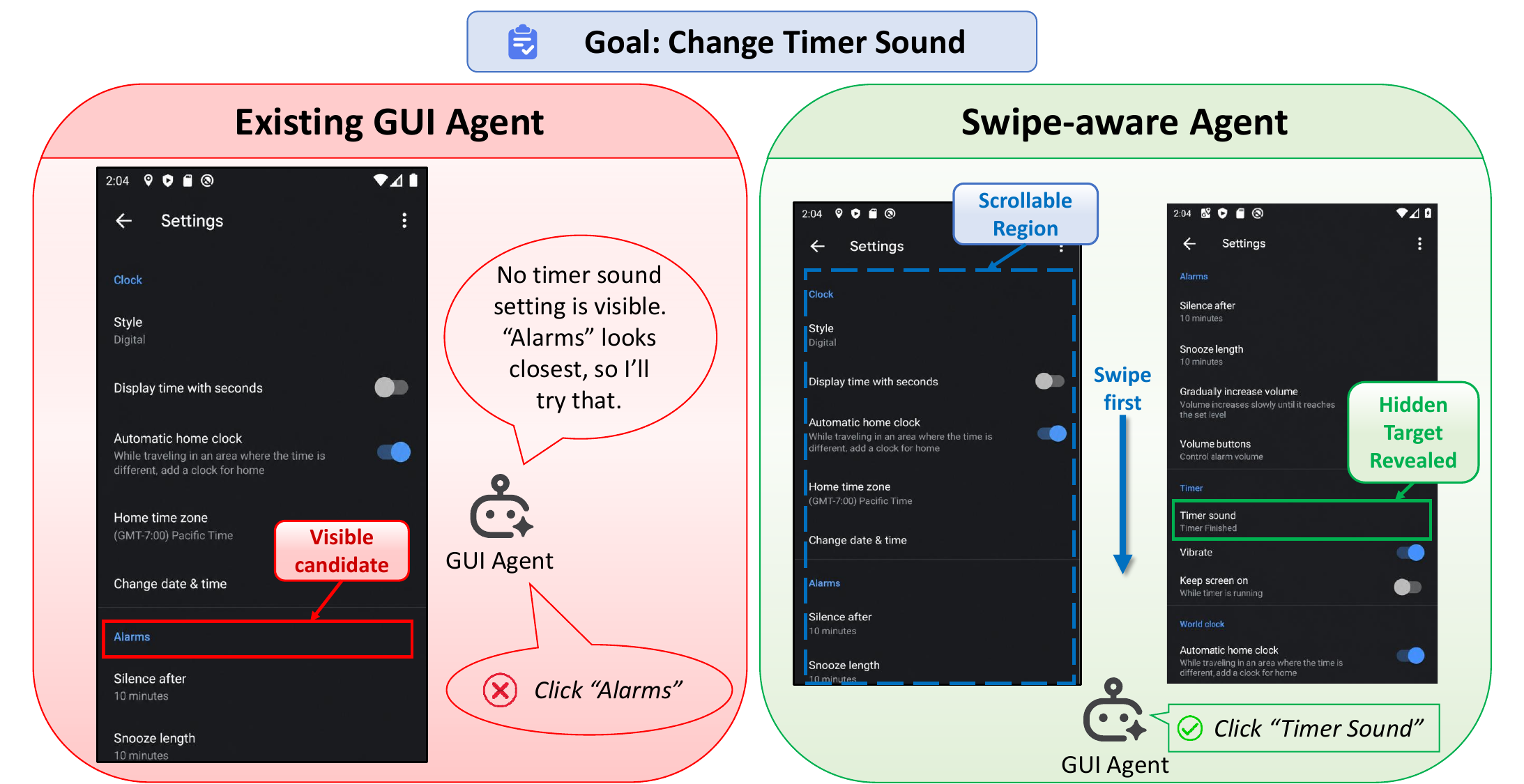}
    \caption{Motivating example of swipe planning for a hidden target.}
    \label{fig:greedy_failure}
\end{figure}

However, as shown in Figure~\ref{fig:greedy_failure}, when the correct target is hidden, even the most relevant visible element may be an incorrect choice.
An agent that continues to select among visible candidates may therefore force a wrong click instead of reaching the required target, as shown in Figure~\ref{fig:greedy_failure}.
The decision must then extend beyond which element to click to whether exploration is needed to reveal the hidden target.

\subsection{Hidden Area Exploration via Swipe Interaction}

As mobile apps offer more functionality, their increasingly dense interfaces often contain more content than the screen can display at once.
Task-relevant targets may be partially clipped or completely hidden within full-page or local scrollable regions, such as lists and carousels~\cite{rawles2023aitw,rawles2025androidworld,xu2024androidlab}.
When the target lies beyond the visible content, the agent must swipe to reveal it and then reassess the available interaction targets.

Planning this exploration requires the agent to decide when, where, and how to swipe before the target becomes visible.
For example, finding a category hidden in a horizontal carousel requires swiping that carousel rather than clicking another visible category.
The agent must infer the region's function and relevance to the task, locate its boundaries, and choose a suitable swipe direction and extent.
Multiple or nested scrollable regions complicate this decision, as swiping the wrong region may reveal content unrelated to the task.

Existing GUI agents already support swipe as an action primitive~\cite{zhang2023appagent,wang2024mobileagent,qin2025uitars}.
However, this support alone does not ensure that an agent can recognize when exploration is necessary or select the appropriate region to explore.
This distinction motivates a quantitative comparison between click planning with visible targets and swipe planning for hidden-area exploration.

\subsection{Preliminary Study of Scrollable Perception}
\label{ssec:2.3_study}

To examine the difficulty of hidden-area exploration compared with visible-target selection, we conduct a preliminary interaction study.
The study compares click and swipe planning by prompting GUI agents to locate a specific UI element, either visible or hidden, and assessing whether they select the correct next interaction based on the current screenshot.

\subsubsection{Study setup.}
Both general-purpose VLM GUI agents and specialized GUI models are selected for this study.
Specifically, general-purpose VLM GUI agents include Qwen3-VL~\cite{bai2025qwen3vltechnicalreport} and GPT-5.4~\cite{openai2026gpt54}.
As for the specialized GUI models, we include GUI-Owl~\cite{ye2025mobileagentv3}, UI-TARS~\cite{qin2025uitars}, GUI-G\textsuperscript{2}~\cite{tang2025guig2gaussianrewardmodeling}, UI-Venus 1.5~\cite{venusteam2026uivenus15technicalreport} and MAI-UI~\cite{zhou2025maiuitechnicalreportrealworld}.
To control for differences in input, all agents receive the same current screenshot and target element description.

\subsubsection{Task configuration.}

Using an Android phone connected through ADB, we manually collected 218 UI interactions from 55 applications, including 36 Chinese-language and 19 English-language applications.
These interactions include 192 swipe cases requiring exploration to reveal or reach the target component and 26 click cases with visible targets as the reference condition.

Since scrollable regions vary in appearance, we categorize these swipe interactions by direction (vertical or horizontal), scope (full-screen or local), and target visibility (partially visible or completely invisible).
Specifically, full-screen scrolling refers to scrollable regions occupying more than 75\% of the screen, while all remaining cases are classified as local swiping.
As for visibility, we classify a target as completely invisible when less than 25\% of its area is visible, and as partially visible when 25\% to 75\% is visible.
The upper bound reflects the likelihood that targets with more than 75\% of their area visible can already be interacted with directly.

\subsubsection{Evaluation metrics.}
Instead of running each agent end-to-end on complete tasks, we focus on whether it can produce the \emph{correct next interaction} to facilitate comparison across agents.
Although swipe actions involve multiple parameters, including position, direction, and distance~\cite{rawles2023aitw}, this study focuses on position and direction, since the remaining parameters are less relevant to scrollable perception.
We then report the proportion of correctly planned interactions in each task category to compare click and swipe planning.

\begin{table*}[t]
\centering
\scriptsize
\caption{Success rates (\%) of representative GUI agents on click and swipe step-level interaction tasks. Part-Inv. and Inv. denote partially visible and completely invisible targets, respectively.}
\label{tab:gui_agent_swipe_motivation}
\begin{threeparttable}
\setlength{\tabcolsep}{2pt}
\renewcommand{\arraystretch}{1.25}
\renewcommand{\tabularxcolumn}[1]{m{#1}}

\begin{tabularx}{\textwidth}{@{}*{10}{>{\centering\arraybackslash}X}@{}}
\toprule

\multirow[c]{2}{=}[-0.4ex]{\centering\textbf{Action}}
& \multirow[c]{2}{=}[-0.4ex]{\centering\textbf{Scope}}
& \multirow[c]{2}{=}[-0.4ex]{\centering\textbf{Visibility}}
& \multicolumn{2}{c}{\textbf{General-purpose Models}}
& \multicolumn{5}{c}{\textbf{Specialized GUI Models}} \\
\cmidrule(lr){4-5} \cmidrule(lr){6-10}

& & 
& \makebox[\linewidth][c]{\textbf{Qwen3-VL}}
& \textbf{GPT-5.4} 
& \textbf{GUI-Owl} 
& \textbf{UI-TARS} 
& \textbf{GUI-G\textsuperscript{2}}
& \textbf{UI-Venus 1.5} 
& \textbf{MAI-UI} \\
\midrule

Click & -- & Visible 
& 96.2 & 96.2 
& 100.0 & 50.0 & 88.5 & 100.0 & 100.0 \\
\midrule

\multirow[c]{4}{=}{\centering Vertical Swipe}
 & \multirow[c]{2}{=}{\centering Full-screen}
 & Part-Inv. 
 & 71.4 & 47.6 
 & 61.9 & 71.4 & 66.7 & 66.7 & 52.4 \\
 &  & Inv. 
 & 76.5 & 61.8 
 & 61.8 & 64.7 & 47.1 & 64.7 & 61.8 \\
\cmidrule(lr){2-10}

 & \multirow[c]{2}{=}{\centering Local}
 & Part-Inv. 
 & 42.9 & 47.6 
 & 47.6 & 52.4 & 38.1 & 33.3 & 42.9 \\
 &  & Inv. 
 & 62.1 & 69.0 
 & 58.6 & 48.3 & 55.2 & 51.7 & 58.6 \\

\midrule

\multirow[c]{4}{=}{\centering Horizontal Swipe}
 & \multirow[c]{2}{=}{\centering Full-screen}
 & Part-Inv. 
 & 11.8 & 11.8 
 & 11.8 & 5.9 & 29.4 & 5.9 & 17.6 \\
 &  & Inv. 
 & 17.6 & 23.5 
 & 17.6 & 23.5 & 23.5 & 5.9 & 11.8 \\
\cmidrule(lr){2-10}

 & \multirow[c]{2}{=}{\centering Local}
 & Part-Inv. 
 & 23.8 & 38.1 
 & 19.0 & 38.1 & 19.0 & 4.8 & 19.0 \\
 &  & Inv. 
 & 37.5 & 59.4 
 & 34.3 & 28.1 & 43.8 & 21.9 & 34.4 \\

\bottomrule
\end{tabularx}
\end{threeparttable}
\end{table*}

\subsubsection{Results and Implications.}

Table~\ref{tab:gui_agent_swipe_motivation} shows a substantial gap between visible-target clicking and hidden-area exploration.
Weighting the eight swipe categories by their actual sample counts and averaging across the evaluated agents, the mean success rate is 90.1\% for click cases and 42.9\% for swipe cases, a decrease of 47.3 percentage points.
Moreover, agents' strong click capabilities do not translate into reliable swipe planning. For example, GUI-Owl, UI-Venus 1.5, and MAI-UI each achieve 100\% on click cases, but their sample-weighted swipe success rates are only 42.2\%, 35.4\%, and 40.6\%, respectively.

Direction is the clearest source of variation: performance averages 57.4\% for vertical swipes but only 25.3\% for horizontal swipes.
This pattern is consistent with the vertical scrolling convention used for UI lists~\cite{materialdesignlists}.
Moreover, target visibility is associated with a substantial difference in planning accuracy: completely invisible targets achieve a higher weighted average than partially visible targets (47.2\% versus 36.8\%).
The gap between full-screen and local swiping is the smallest (44.6\% versus 41.3\%), suggesting that region size is not the primary bottleneck in current agents' GUI understanding.

Further analysis reveals a dominant failure mode: in 83.73\% of failed swipe tasks, agents decided to click on an incorrect visible target instead of exploring the correct but hidden region.
Only fewer than 30\% of failures involve selecting the wrong scrollable region or swipe direction.
These findings reveal a strong tendency for LLM-based GUI agents to settle for seemingly suitable visible components rather than explore hidden areas.

These results indicate that merely including swipe in an agent's action space is far from sufficient for reliable GUI interaction.
Agents still struggle to determine whether exploration is necessary and, if so, where and in which direction to swipe.
These findings motivate our subsequent investigation into prompting agents with scrollable-region information to encourage exploration and reduce incorrect clicks.

\section{Method}
\label{sec:method}

This section presents \MIG, a scrollable-region perception module that supports hidden-area exploration through non-intrusive augmentation of GUI agents' screenshots and prompts.

\subsection{Overview}

\begin{figure*}[t]
    \centering
    \includegraphics[width=\textwidth]{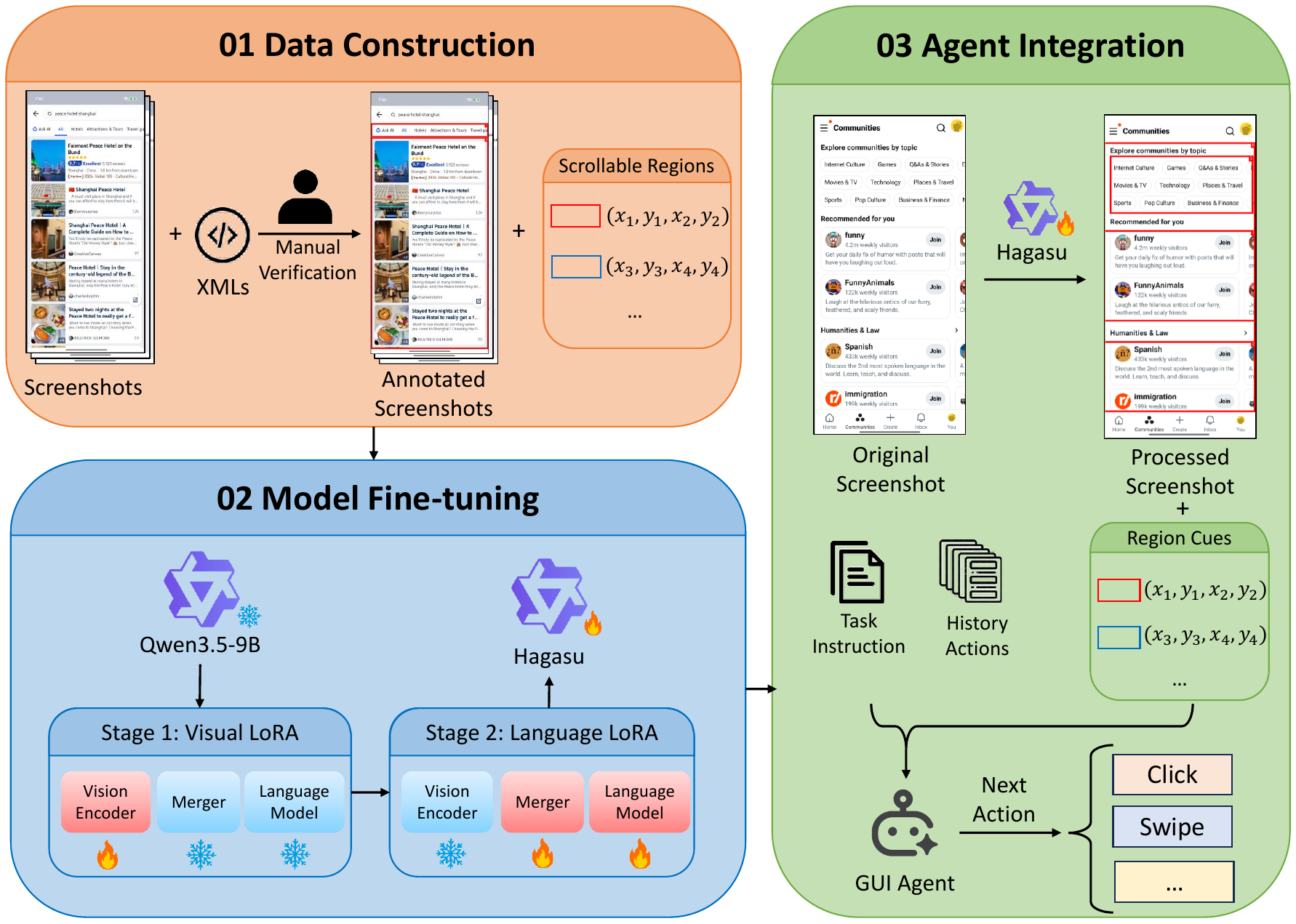}
    \caption{Overview of \MIG.}
    \label{fig:workflow}
\end{figure*}

As shown in Figure~\ref{fig:workflow}, the \MIG workflow consists of dataset construction, model fine-tuning, and agent integration.
First, we collect Android screenshots and accessibility XML through ADB and manually verify and refine XML-derived region candidates to construct screenshot--region pairs.
Second, we build on Qwen3.5-9B's UI understanding capabilities through two-stage LoRA fine-tuning, enabling it to predict all scrollable regions as a JSON list of bounding boxes.
Third, when integrated into a GUI agent, \MIG marks its predicted regions on the screenshot at each step and supplements the agent's prompt with cues for swipe-based exploration.
The base agent combines these cues with the task and interaction history to plan its next action, retaining control over whether and how to click or swipe.

\subsection{Dataset Construction}

\subsubsection{Data Acquisition and Scrollable-Region Annotation.}

Although public GUI interaction datasets such as Android in the Wild (AITW)~\cite{rawles2023aitw} and GUIOdyssey~\cite{lu2025guiodyssey} provide screenshots and action trajectories, they lack complete bounding-box annotations of all scrollable regions in each screenshot.
As shown in Table~\ref{tab:rq1_swipe_perception}, even existing first-tier models struggle to identify scrollable regions accurately, making automatic annotation infeasible.
We therefore manually collect and annotate UI screenshots to provide reliable supervision for scrollable-region perception training.

49 apps are selected from Google Play's popular app rankings~\cite{googleplayranking}, including 33 Chinese- and 16 English-language apps.
For each app, we manually navigate its interface and save screenshots containing scrollable regions.
To reduce annotation effort, we also capture the corresponding accessibility XML for each screenshot through Android Debug Bridge (ADB)~\cite{androidadb}, since the UI component attributes in the XML can support preliminary region annotation~\cite{androidxmlhierarchy}.
Specifically, we parse each XML node’s bounds attribute into a screen-coordinate rectangle, retain nodes whose scrollable attribute is set to true and valid bounds cover more than 4\% of the screen, deduplicate and clip these boxes, and combine them with image-based proposals to form the candidate set for manual verification.
Note that the XML serves only as an annotation aid and is not required by \MIG at inference time.

Finally, three authors of this paper manually annotate the scrollable regions using these candidates as a starting point.
Specifically, two authors independently inspect each screenshot, retain valid candidates, remove incorrect ones, and draw boxes for missed regions.
The third author resolves disagreements to finalize the annotations.
The annotations include each region's bounding box and swipe direction.
In particular, for screenshots with multiple scrollable regions, we annotate every region, including full-page containers and local components such as lists and carousels.

\subsubsection{Dataset Organization and Split.}

\begin{figure}[t]
    \centering
    \subfigure[]{\includegraphics[width=0.4\columnwidth]{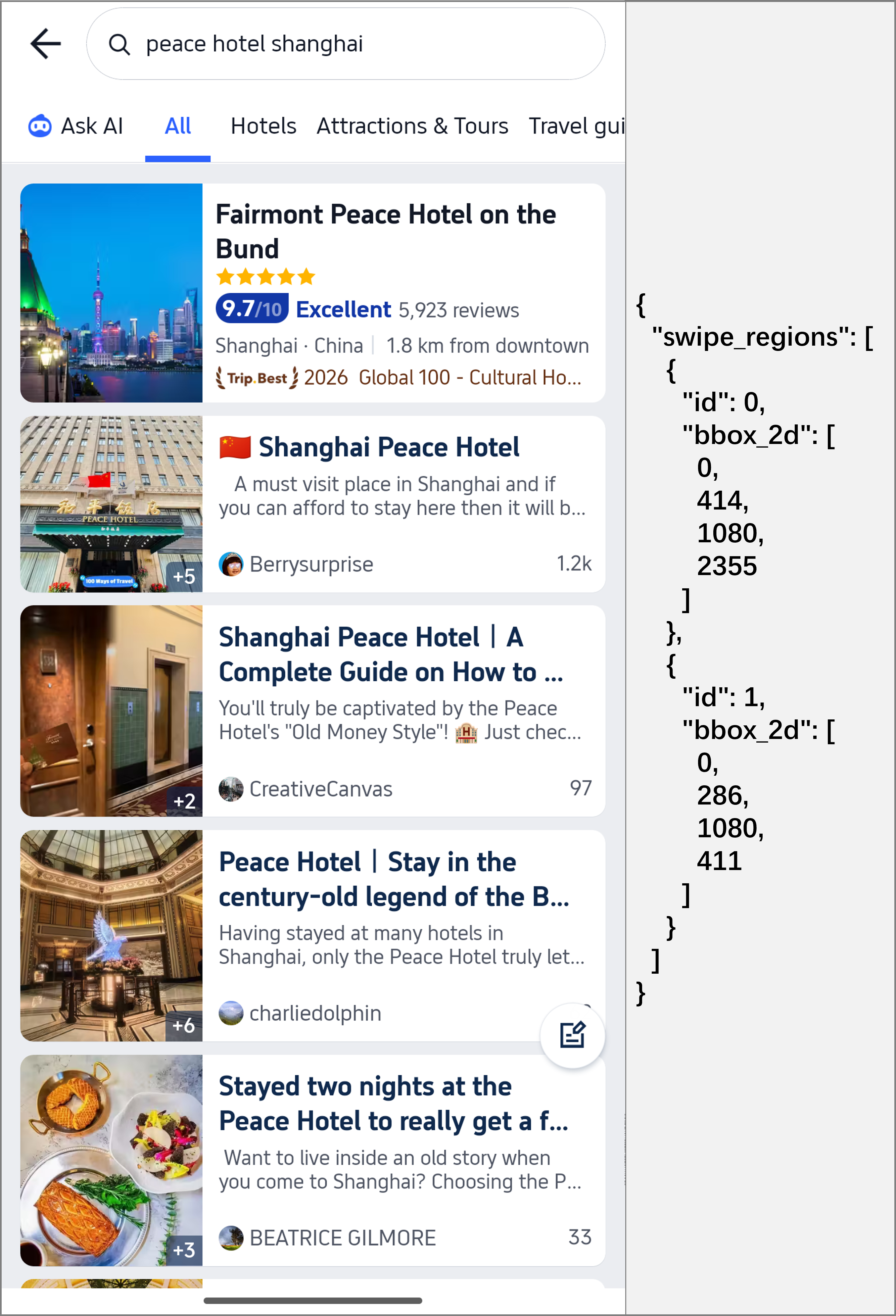}}
    \hspace{0.02\columnwidth}
    \subfigure[]{\includegraphics[width=0.4\columnwidth]{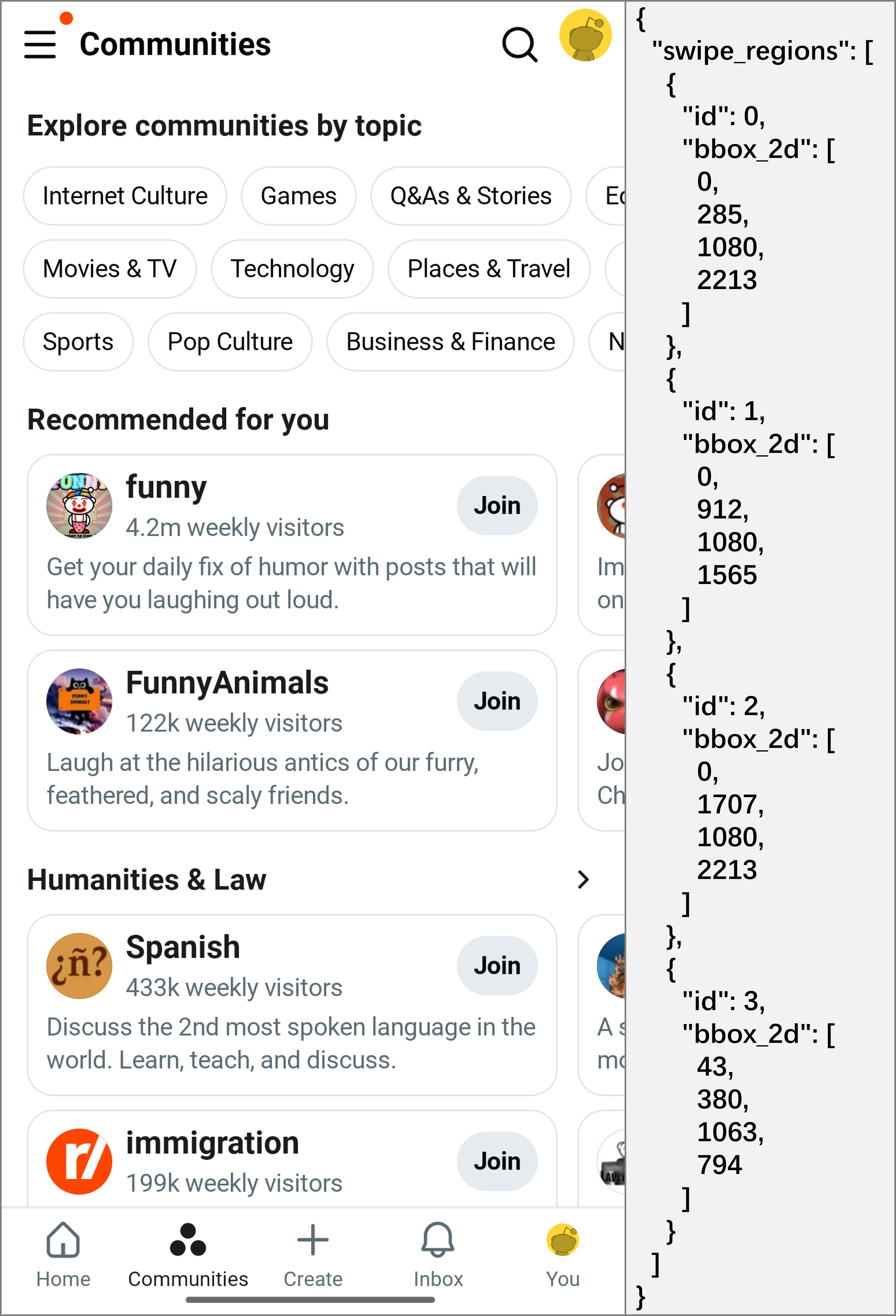}}
    \caption{Examples of screenshot--JSON samples in the scrollable-region dataset. Each JSON lists the complete set of annotated regions in its corresponding screenshot, with coordinates following the \texttt{[x\_min, y\_min, x\_max, y\_max]} convention.}
    \label{fig:data_example}
\end{figure}

As illustrated in Figure~\ref{fig:data_example}, each screenshot is paired with a JSON file containing the complete set of annotated bounding boxes.
The dataset is represented as:
\begin{equation}
    \mathcal{D} = \{(I_i, B_i)\}_{i=1}^{N}, \qquad
    B_i = \{b_{i1}, b_{i2}, \ldots, b_{im_i}\},
\end{equation}
where $N$ is the number of samples, $I_i$ is the $i$-th screenshot, $m_i$ is its number of scrollable regions, and $b_{ij}$ is the bounding box of the $j$-th region.
We will further train \MIG to predict the complete region set $B_i$ for each screenshot.

The resulting dataset contains 667 screenshots, including 364 with a single annotated region and 303 with multiple regions.
Following the principle of separating model training, model selection, and final evaluation~\cite{goodfellow2016deep}, we randomly divide the dataset into 540 training, 60 validation, and 67 test samples.

\subsection{Model Fine-tuning}

Since general-purpose multimodal LLMs struggle to accurately identify scrollable regions, as shown in Table~\ref{tab:rq1_swipe_perception}, we fine-tune a pretrained vision-language model on our manually collected and annotated screenshots to specialize it for scrollable-region perception.
The input consists of a screenshot $I_i$ and a fixed textual prompt asking the model to identify all scrollable regions, without requiring accessibility XML or a task-specific target.
The model is supervised with the annotated region set $B_i$ and produces a JSON object containing a variable-length list of predicted bounding boxes:
\begin{equation}
    \hat{B}_i = f_\theta(I_i, p) = \{\hat{b}_{ij}\}_{j=1}^{\hat{m}_i},
\end{equation}
where $f_\theta$ denotes the fine-tuned model with parameters $\theta$, $p$ is the fixed textual prompt, $\hat{m}_i$ is the number of predicted regions, and $\hat{b}_{ij}$ is the bounding box of the $j$-th predicted region.

\subsubsection{Multi-Region Sampling.}

During annotation, we found that many UI screens contain multiple scrollable regions.
Identifying the complete region set on these screens is substantially more challenging than detecting a single region, as local scrollable regions can easily be missed.
Specifically, 303 of the 667 screenshots contain multiple scrollable regions, accounting for 45.4\% of the we selected screenshots.
Thus, these screens reflect a common pattern in practical mobile interfaces, where a full-page container may coexist with local lists, carousels, or other scrollable components.
All scrollable regions on these screens must be identified, since the task-relevant component may be located in any one of them.

To address low recall on screens with multiple scrollable regions, we therefore oversample training screenshots according to the number of annotated regions.
Specifically, the number of repetitions per epoch for each training screenshot is equal to its number of annotated scrollable regions, capped at five.
Table~\ref{tab:rq1_swipe_perception} in the following Section~\ref{sec:evaluation} shows that this reweighting effectively improves the model's recall by increasing its exposure to multi-region examples during training.

\subsubsection{Two-Stage LoRA SFT}

Scrollable regions lack clear visual indicators, making it difficult for models to recognize them and associate their visual appearance with swipe interaction semantics.
Given the limited size of our manually collected dataset, we adopt LoRA~\cite{hu2022lora} to adapt the model through a small number of trainable parameters instead of full-parameter fine-tuning.
To address this challenge while making effective use of the limited data, we organize LoRA fine-tuning into two stages.
The first stage strengthens visual perception of UI images, focusing on the appearance and spatial structure of UI regions.
The second stage fine-tunes the language model to connect swipe interaction semantics with the visual features learned in the first stage.

Following prior work on Qwen-based GUI models~\cite{cheng2024seeclick,lin2024showui}, we select Qwen3.5-9B as the backbone of \MIG.
Specifically, the first stage trains LoRA adapters in the visual component while keeping the original visual weights, the merger, and the language model frozen.
The trained visual adapters are then merged into the backbone to initialize the second stage.
In the second stage, we train newly introduced language LoRA adapters and all merger parameters while keeping the adapted visual component and the original language-model weights frozen.
Both stages use LoRA with rank 16, $\alpha=32$, and dropout 0.1, with supervision provided by screenshots paired with JSON annotations and training samples drawn using the multi-region oversampling strategy described above.

The 667 screenshots are randomly divided into 540 training, 60 validation, and 67 test samples.
After oversampling, the 540 training screenshots yield 964 training entries per epoch, while the validation and test sets remain unchanged.
The validation set is used for model selection, and the test set is reserved for evaluation in Section~\ref{sec:evaluation}.

\subsection{Agent Integration}
\label{sec:agent_integration}

\subsubsection{Screenshot Annotation and Interaction Loop.}
We make minimal modifications to integrate \MIG into various GUI agents as a non-intrusive plug-in, providing supplementary visual and textual cues for the model's automated exploration.

\begin{figure}[t]
	\centering
	\subfigure[]{\includegraphics[width=0.30\columnwidth]{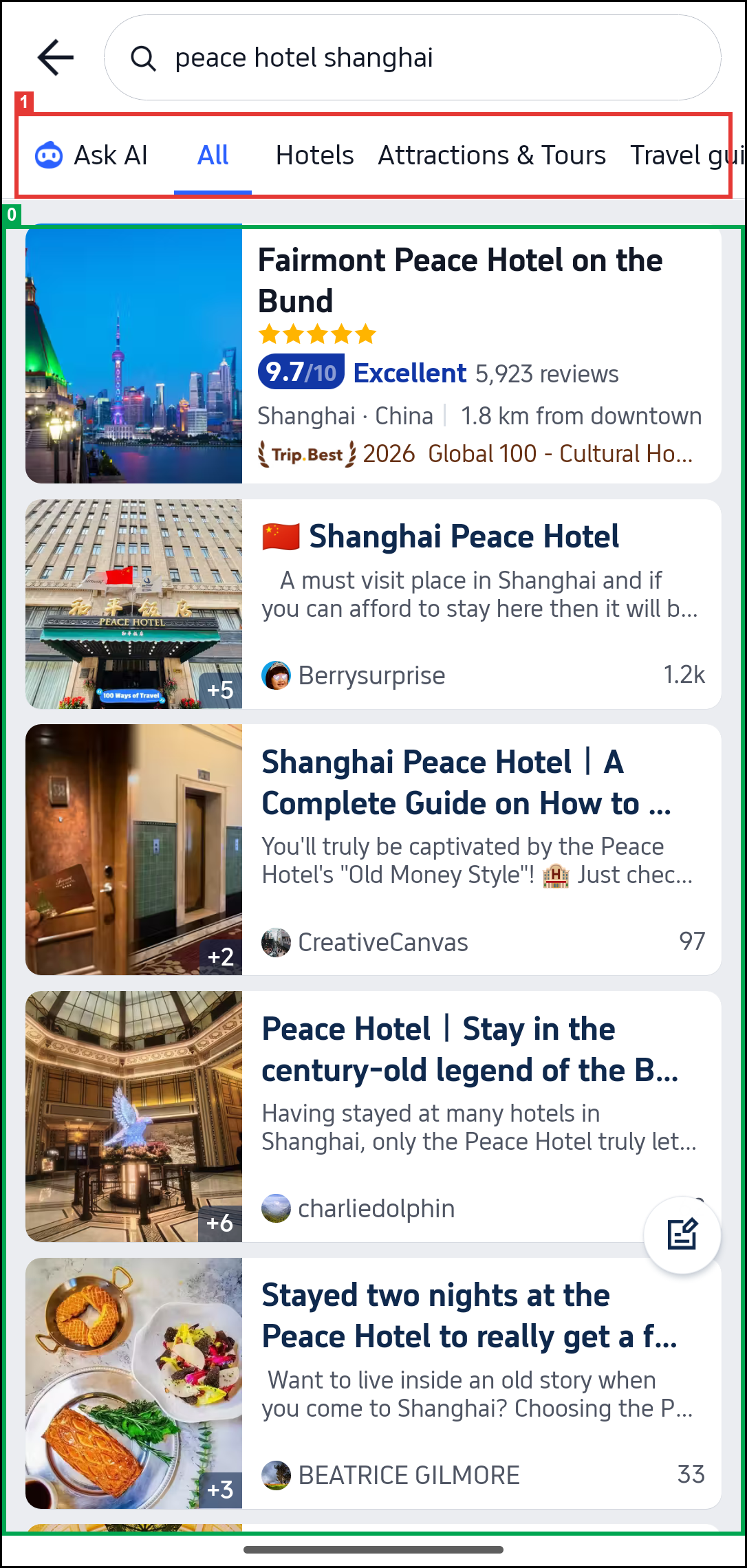}}
	\hspace{0.02\columnwidth}
	\subfigure[]{\includegraphics[width=0.30\columnwidth]{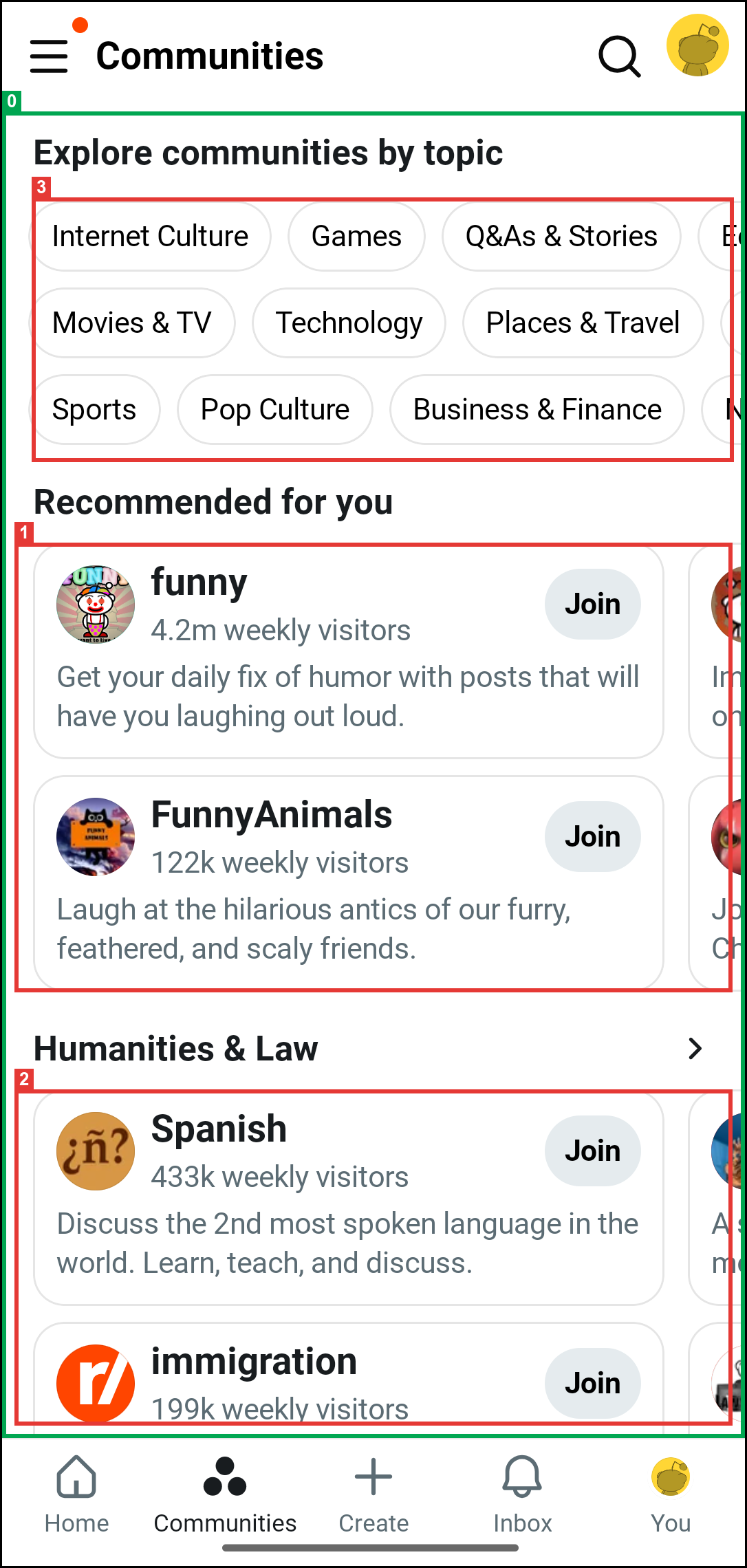}}
	\caption{Scrollable-region annotations predicted by \MIG on the screenshots from Figure~\ref{fig:data_example}. Green and red boxes indicate vertically and horizontally scrollable regions, respectively, with region IDs at the top-left corners.}
	\label{fig:annotation_example}
\end{figure}

At each interaction step, agent captures a screenshot and passes it to \MIG, which returns a JSON list of bounding boxes for the predicted scrollable regions.
As shown in Figure~\ref{fig:annotation_example}, we annotate the screenshot with numbered boxes and add the following cue $c$ to the agent's prompt: \emph{``The screenshot has numbered boxes marking predicted scrollable regions.''}
Using the task instruction, annotated screenshot, and interaction history, the agent generates its next action, executes it on the device, and captures a new screenshot for subsequent planning. Algorithm~\ref{alg:agent_execution} provides a pseudocode summary of our agent loop.

\begin{algorithm}[t]
	\caption{GUI Agent Execution with \MIG}
	\label{alg:agent_execution}
	\KwIn{Task $T$, agent prompt $P$, region cue $c$, maximum device actions $\mathit{budget}$}
	$P_{\mathrm{aug}} \leftarrow \mathrm{SupplementPrompt}(P,c)$\;
	$\mathit{hist} \leftarrow []$; $\mathit{step} \leftarrow 0$\;
	\While{true}{
		$\mathit{S} \leftarrow \mathrm{Screenshot}()$\;
		$\hat{B} \leftarrow \text{\MIG}(\mathit{S})$\;
		$\mathit{S}_\mathit{anno} \leftarrow \mathrm{DrawBoxes}(\mathit{S},\hat{B})$\;
		$\mathit{act} \leftarrow \mathrm{AgentDecide}(
		T,\mathit{S}_\mathit{anno},P_{\mathrm{aug}},\mathit{hist},\mathit{step},\mathit{budget})$\;
		\If{$\mathit{act} = \mathrm{finish}$}{
			\textbf{return} AwaitHumanJudgment\;
		}
		\If{$\mathit{step} \ge \mathit{budget}$}{
			\textbf{return} BudgetExhausted\;
		}
		$\mathrm{ExecuteOnDevice}(\mathit{act})$\;
		$\mathit{step} \leftarrow \mathit{step}+1$\;
		$\mathit{S}' \leftarrow \mathrm{Screenshot}()$\;
		$\mathit{mem} \leftarrow \mathrm{AgentMemory}( \mathit{S},\mathit{S}',\mathit{act})$\;
		$\mathit{hist} \leftarrow \mathit{hist} \mathbin{\|} [\mathit{mem}]$\;
	}
\end{algorithm}

When integrated into GUI agents, \MIG is called for every screenshot.
Although \MIG modifies the screenshot and prompt, these modifications serve only as supplementary information about where swipe exploration is possible.
They neither force the agent to make any particular decision nor override its action policy.
Therefore, these agents retain full control over whether and how to click, swipe, or perform other supported actions based on the task discription and the current interaction state.

\subsubsection{Memory Management.}
Although this paper simplifies most GUI-agent components, such as reflection, to highlight the effect of \MIG on planning, memory remains essential for multi-step interaction tasks.
We therefore provide a simplified memory management module to support subsequent evaluation.
Specifically, two memory management strategies are designed for offline and online experiments.
In offline experiments, the history briefly describes the screenshots and actions from preceding steps.
Since the preceding steps are deterministic, we prepared a predetermined interaction history for each step.
At each step, the model receives the interaction history of all preceding steps in that task.
In online experiments, the agents will generate their own interaction history after each step while performing the task.
In addition to screenshot and action descriptions in the same format as in offline experiments, the agent also retains the previous screenshot and compares it with the current one to determine whether the action changed the page, helping reduce the risk of entering an infinite loop.

\section{Experiment Design}
\label{sec:eva_design}

Experiments are conducted to evaluate \MIG's swipe perception capability, its support for swipe planning, and its impact on end-to-end task completion when integrated into GUI agents. The three research questions are as follows.

\begin{itemize}[leftmargin=*]
    \item \textbf{RQ1: Scrollable Region Perception.} How accurately can \MIG detect and localize scrollable regions in mobile UIs?
    \item \textbf{RQ2: Swipe-aware Planning.} How effectively does scrollable-region perception help GUI agents make correct swipe decisions?
    \item \textbf{RQ3: Agent Integration.} How much does \MIG affect GUI agents' end-to-end task completion on swipe-dependent tasks, and what overhead does it introduce?
\end{itemize}

\subsection{Datasets}

For \textbf{RQ1}, we evaluate whether \MIG can correctly identify all scrollable regions in a UI screenshot and accurately localize their boundaries.
We use the dataset described in Section~\ref{sec:method}, containing 67 mobile UI screenshots.
Each screenshot is annotated with all scrollable regions and their boundaries.
Note that a screenshot may contain multiple scrollable regions. 
Therefore, we compare the predicted region set with the complete ground-truth set.

For \textbf{RQ2}, we construct an offline benchmark from human-recorded successful trajectories of 60 mobile GUI interaction tasks, with each trajectory containing an average of 4.1 actions and at least one swipe.
Specifically, this benchmark contains 245 interaction steps, including 99 swipes (39 vertical and 60 horizontal) and 146 clicks.
It evaluates whether agents can correctly decide between acting on visible content and exploring off-screen content when prompted with \MIG's scrollable-region predictions.

For \textbf{RQ3}, considering execution cost and time overhead, we randomly select 18 swipe-dependent tasks across 16 Android apps from the same task pool as RQ2 for end-to-end online evaluation.
Each task requires at least one swipe to reveal off-screen task-relevant content, allowing us to examine whether improved swipe planning translates into higher task completion rates.
The tasks cover content discovery, media playback, product search, and travel-related information retrieval.

\subsection{Baselines}
For \textbf{RQ1}, we compare \MIG with both untuned base models and several training variants to understand where the improvement in scrollable-region perception comes from.
Specifically, we first evaluate two untuned baselines, \textit{Qwen3-VL-8B}~\cite{bai2025qwen3vltechnicalreport} and \textit{Qwen3.5-9B}~\cite{qwen2026qwen35_9b}.
We also include \textit{Qwen3.5-9B + XML}, where scrollable-region candidates extracted from XML are provided to the model for visual refinement.
This comparison examines whether explicit structural cues can compensate for the lack of task-specific fine-tuning.

\begin{figure*}[t]
    \centering
    \includegraphics[width=\textwidth]{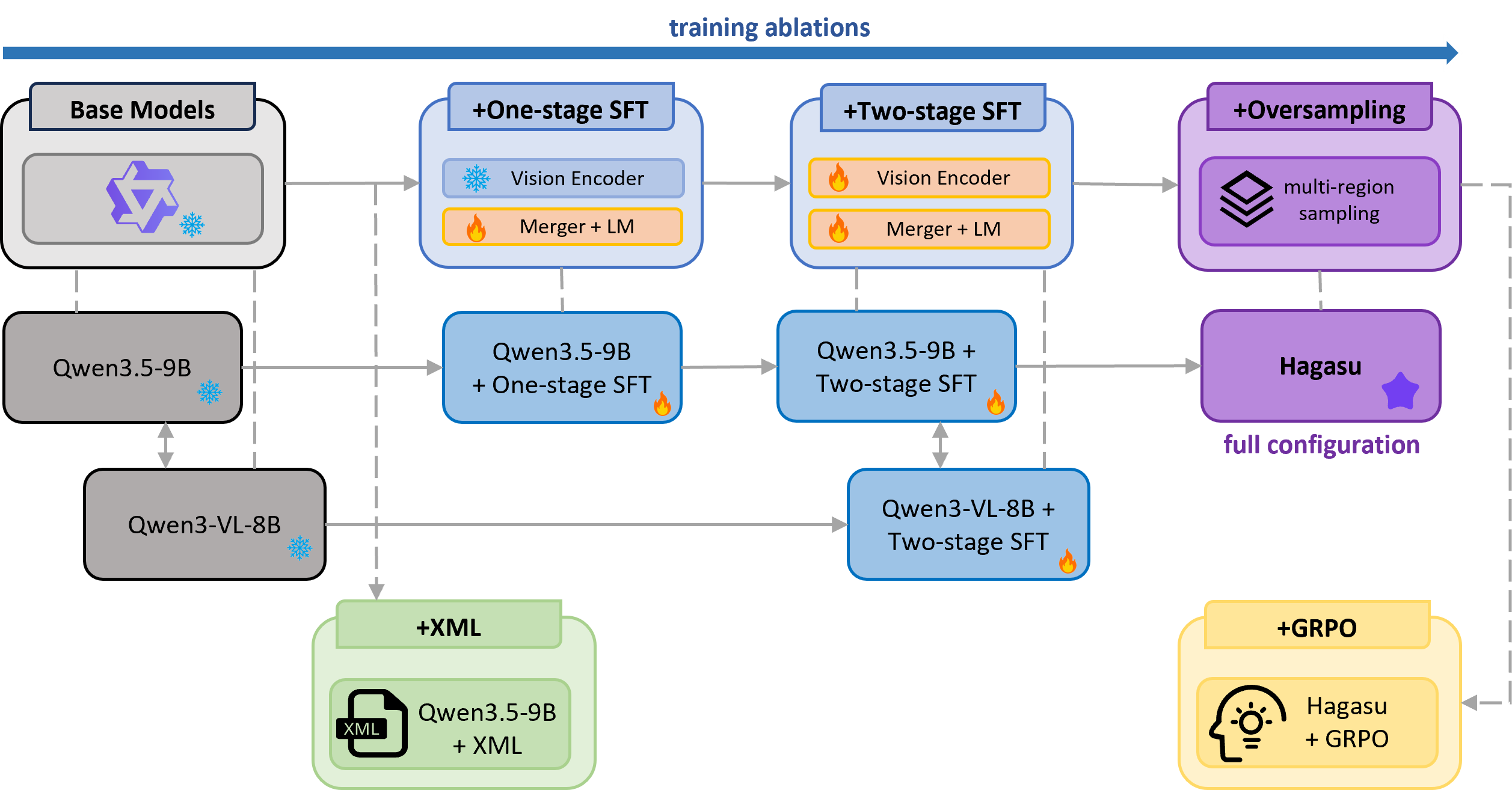}
    \caption{Baseline and ablation configurations for RQ1. Arrows indicate the controlled change from one configuration to another.}
    \label{fig:baseline_ablation}
\end{figure*}

We then construct several variants of \MIG to study its training design.
\textit{Qwen3.5-9B + one-stage SFT} directly fine-tunes the merger and language-model layers while keeping the visual encoder frozen.
\textit{Qwen3.5-9B + two-stage SFT} further introduces the visual fine-tuning stage described in Section~\ref{sec:method}.
Besides, to examine the effect of the base model, we apply the same two-stage SFT procedure to Qwen3-VL-8B, yielding \textit{Qwen3-VL-8B + two-stage SFT}.
Our full \MIG configuration further applies multi-region oversampling on top of Qwen3.5-9B with two-stage SFT.
Finally, we append a GRPO stage to the full configuration to examine whether reinforcement learning provides additional gains.
The relationship among these configurations is summarized in Figure~\ref{fig:baseline_ablation}.

For \textbf{RQ2}, we evaluate nine VLMs with different model scales and GUI capabilities: GPT-6-Luna~\cite{openai2026gpt6luna}, GPT-5.6-Luna~\cite{openai2026gpt56luna}, Seed-2.1-Pro~\cite{bytedance2026seed21}, Seed-Evolving~\cite{volcengine2026seedevolving}, Qwen3.8-27B~\cite{qwen2026qwen38_27b}, Qwen3.7-Plus~\cite{qwen2026qwen37plus}, Qwen3.5-9B~\cite{qwen2026qwen35_9b}, DeepSeek-V4-Flash-Vision~\cite{deepseek2026v4flashvision}, and GUI-OWL-7B~\cite{ye2025mobileagentv3}.

Each model is evaluated under three conditions:
\textit{V} stands for Vanilla, which uses the original screenshot as input.
+\MIG is abbreviated as \textit{H}, scrollable regions predicted by \MIG are highlighted on the input screenshots.
We further include a \textit{G}(+Ground Truth) condition, where manually annotated scrollable regions are highlighted. 
This condition approximates an upper bound for region-aware planning and helps separate errors caused by \MIG.

For \textbf{RQ3}, we select three strong-performing VLMs from RQ2 for end-to-end evaluation: Qwen3.8-27B, Qwen3.7-Plus, and GPT-6-Luna.
Each model is evaluated in both \textit{V}(Vanilla) and \textit{H}(+\MIG) settings.
We repeat each task three times under each model and setting to account for variability in online execution.

\subsection{Metrics}
For \textbf{RQ1}, we evaluate two aspects of scrollable-region perception: whether \MIG identifies the correct set of regions and how accurately it localizes their boundaries.

To evaluate \textit{region detection capability}, we report Precision, Recall, F1, and Exact Match.
Predicted and ground-truth regions are matched one-to-one within each screenshot using the Hungarian algorithm~\cite{kuhn1955hungarian}, with IoU $\geq 0.5$ as the matching threshold.
Matched predictions are counted as true positives, while unmatched predictions and ground-truth regions are counted as false positives and false negatives, respectively.
Precision, Recall, and F1 are micro-averaged over the entire test set.
Exact Match measures the percentage of screenshots for which the complete set of scrollable regions is correctly detected, with neither missing nor extra regions.

As for \textit{localization accuracy}, we report set-level mIoU.
For each screenshot, we perform one-to-one matching between predicted and ground-truth regions to maximize their total IoU.
The summed IoU is normalized by the larger number of predicted or ground-truth regions, such that both missing and extra regions are penalized.
We then average the resulting scores across all screenshots.

For \textbf{RQ2}, we report three metrics to evaluate both the benefit and potential side effects of scrollable-region cues.
\textit{Swipe Prediction Recall} is the percentage of the 99 reference swipe steps correctly predicted as swipes, measuring the agent's ability to identify when exploration is needed.
\textit{Click Prediction Recall} is the percentage of the 146 reference click steps correctly predicted as clicks, measuring whether region cues preserve normal click decisions.
Finally, \textbf{Click$\rightarrow$Swipe Rate} is the percentage of reference click steps incorrectly predicted as swipes, specifically measuring whether the cues cause agents to over-select swipe actions.

For \textbf{RQ3}, we report task success rate (SR) and $\sigma$(SR) to assess task completion and stability, with $\sigma$(SR) denoting the population standard deviation of the three round-level success rates.
We compare the average number of interaction steps across all runs and across successful runs in each setting to evaluate interaction efficiency.
We also report average execution time (Avg. Time) and input/output token usage (Avg. I Token and Avg. O Token) per run to measure execution cost.
Token usage covers reported agent-model decision and memory generation calls and \MIG region-annotation inference, with each component's mean rounded down to an integer.

\subsection{Implementation}

We collect mobile UI screens from Android devices through ADB~\cite{androidadb}, recording both the screenshot and its accessibility XML for each UI state.
During annotation, XML- and image-based proposals are used only as candidate regions to reduce manual effort.
Two annotators independently verify these candidates, remove invalid regions, and add missed regions, while disagreements are resolved by a third annotator.
The final annotation contains the complete set of manually verified scrollable-region bounding boxes for each screenshot.

\MIG is fine-tuned from a vision-language backbone using screenshot--region supervision.
Given a screenshot, the model outputs the bounding boxes of all detected scrollable regions in JSON format.

In the plug-in setting, an annotation service retrieves or predicts region boxes for each decision screenshot, renders them on the image, and adds a textual cue that the marked regions are scrollable. The base agent then combines this augmented observation with the task and interaction history to decide whether and how to click or swipe, while retaining control of action execution.

For \textbf{RQ3}, tasks are executed on a Xiaomi 13 smartphone running Android 16 with a screen resolution of $2400 \times 1080$ pixels. The initial state is manually restored before each run. For a given task, the instruction and action budget remain fixed across agents, settings, and repetitions.
To reduce repeated annotation, \MIG employs a persistent cache of unannotated screenshots and their predicted region boxes. When the current screenshot matches a cached image of the same dimensions with a structural similarity index (SSIM) of at least 0.90, the boxes from the most similar image are rendered on the current screenshot. Otherwise, \MIG predicts the regions and stores the result for subsequent reuse.

\section{Evaluation Result and Analysis}
\label{sec:evaluation}

\subsection{RQ1: Scrollable-Region Detection}

\begin{table*}[t]
\centering
\scriptsize
\caption{Comparison of \MIG and baselines on scrollable-region detection.}
\label{tab:rq1_swipe_perception}
\begin{threeparttable}
\setlength{\tabcolsep}{5pt}
\renewcommand{\arraystretch}{1.25}
\begin{tabular*}{\linewidth}{@{\extracolsep{\fill}}lccccc}
\toprule
\textbf{Method} & \textbf{Precision@0.5} & \textbf{Recall@0.5} & \textbf{F1@0.5} & \textbf{mIoU} & \textbf{Exact Match@0.5} \\
\midrule
Qwen3-VL-8B & 57.63 & 26.15 & 35.98 & 25.55 & 16.42 \\
Qwen3.5-9B & 58.46 & 29.23 & 38.97 & 31.63 & 23.88 \\
Qwen3.5-9B + XML & 62.30 & 52.05 & 56.72 & 54.30 & 44.68 \\
\midrule
Qwen3.5-9B + one-stage SFT & \textbf{78.00} & 60.00 & 67.83 & 54.21 & 41.79 \\
Qwen3.5-9B + two-stage SFT & 73.68 & 64.62 & 68.85 & 56.24 & \textbf{44.78} \\
Qwen3-VL-8B + two-stage SFT & 70.00 & 59.23 & 64.17 & 50.63 & 32.84 \\
\midrule
\MIG & 68.46 & \textbf{78.46} & \textbf{73.12} & \textbf{60.20} & 43.28 \\
\MIG + GRPO & 65.58 & 77.69 & 71.13 & 58.79 & 38.81 \\
\bottomrule
\end{tabular*}
\end{threeparttable}
\end{table*}

\paragraph{Overall performance.}
\MIG substantially outperforms the untuned vision-language models in scrollable-region perception.
As shown in Table~\ref{tab:rq1_swipe_perception}, \MIG achieves 78.46\% Recall and 73.12\% F1, improving over untuned Qwen3.5-9B by 49.23 and 34.15 percentage points, respectively.
Providing Qwen3.5-9B with XML-derived candidate regions improves F1 from 38.97\% to 56.72\%, confirming that structural cues are useful.
Nevertheless, \MIG further improves F1 to 73.12\%, showing that task-specific fine-tuning is substantially more effective than directly relying on XML candidates.

\paragraph{Effect of training designs.}
Multi-region oversampling primarily improves region coverage.
Removing it reduces Recall from 78.46\% to 64.62\% and F1 from 73.12\% to 68.85\%, while increasing Precision from 68.46\% to 73.68\%.
This trade-off is consistent with the goal of oversampling: reducing missed regions at the cost of some additional false positives.
The two-stage SFT schedule provides a smaller but consistent improvement over one-stage SFT, higher Recall, mIoU, and Exact Match.
These results support the use of both visual fine-tuning and multi-region oversampling in \MIG.

\paragraph{Other design choices.}
Appending GRPO to \MIG provides no additional benefit, reducing F1 from 73.12\% to 71.13\% and Exact Match from 43.28\% to 38.81\%.
This suggests that predicting floating-point coordinates for multiple regions creates a large search space that may hinder reinforcement learning convergence with limited training data.
Despite the substantial improvement in region coverage, complete region-set recovery remains challenging, with \MIG achieving an Exact Match of 43.28\%.

\subsection{RQ2: Swipe-aware Planning Improvement}

\begingroup
\newcommand{\rqtwocell}[2]{\begin{tabular}[c]{@{}c@{}}#1\\[-1pt]{\scriptsize\ensuremath{(#2)}}\end{tabular}}
\newcommand{\rqonecell}[1]{\begin{tabular}[c]{@{}c@{}}#1\end{tabular}}
\begin{table*}[t]
\centering
\footnotesize
\caption{Offline experiment results.}
\label{tab:rq2_swipe_planning}
\begin{threeparttable}
\setlength{\tabcolsep}{1.5pt}
\renewcommand{\arraystretch}{1.05}
\begin{tabular*}{\linewidth}{@{\extracolsep{\fill}}lccccccccc}
\toprule
\multicolumn{1}{c}{\multirow[c]{3}{*}{\textbf{Model}}}
& \multicolumn{3}{c}{\textbf{Swipe Prediction Recall(\%) $\uparrow$}}
& \multicolumn{3}{c}{\textbf{Click Prediction Recall(\%) $\uparrow$}}
& \multicolumn{3}{c}{\textbf{Click$\rightarrow$Swipe Rate(\%) $\downarrow$}} \\
\cmidrule(lr){2-4}\cmidrule(lr){5-7}\cmidrule(lr){8-10}

& \multirow[c]{2}{*}{\textbf{V}}
& \textbf{H}
& \textbf{G}
& \multirow[c]{2}{*}{\textbf{V}}
& \textbf{H}
& \textbf{G}
& \multirow[c]{2}{*}{\textbf{V}}
& \textbf{H}
& \textbf{G} \\[-1pt]

&
& {\scriptsize $(\Delta V)$}
& {\scriptsize $(\Delta V;\,\Delta H)$}
&
& {\scriptsize $(\Delta V)$}
& {\scriptsize $(\Delta V;\,\Delta H)$}
&
& {\scriptsize $(\Delta V)$}
& {\scriptsize $(\Delta V;\,\Delta H)$} \\

\midrule
\rqonecell{GPT-6-Luna} & \rqonecell{58.59} & \rqtwocell{65.66}{\textcolor{green!50!black}{+7.07}} & \rqtwocell{64.65}{\textcolor{green!50!black}{+6.06};\,\textcolor{red}{-1.01}} & \rqonecell{94.52} & \rqtwocell{94.52}{0.00} & \rqtwocell{91.78}{\textcolor{red}{-2.74};\,\textcolor{red}{-2.74}} & \rqonecell{1.37} & \rqtwocell{2.05}{\textcolor{red}{+0.68}} & \rqtwocell{2.05}{\textcolor{red}{+0.68};\,0.00} \\
\addlinespace[2pt]
\rqonecell{GPT-5.6-Luna} & \rqonecell{50.51} & \rqtwocell{52.53}{\textcolor{green!50!black}{+2.02}} & \rqtwocell{55.56}{\textcolor{green!50!black}{+5.05};\,\textcolor{green!50!black}{+3.03}} & \rqonecell{93.84} & \rqtwocell{91.78}{\textcolor{red}{-2.05}} & \rqtwocell{91.10}{\textcolor{red}{-2.74};\,\textcolor{red}{-0.68}} & \rqonecell{2.74} & \rqtwocell{5.48}{\textcolor{red}{+2.74}} & \rqtwocell{4.79}{\textcolor{red}{+2.05};\,\textcolor{green!50!black}{-0.68}} \\
\addlinespace[2pt]
\rqonecell{Seed-2.1-Pro} & \rqonecell{51.52} & \rqtwocell{55.56}{\textcolor{green!50!black}{+4.04}} & \rqtwocell{53.54}{\textcolor{green!50!black}{+2.02};\,\textcolor{red}{-2.02}} & \rqonecell{92.47} & \rqtwocell{93.84}{\textcolor{green!50!black}{+1.37}} & \rqtwocell{91.78}{\textcolor{red}{-0.68};\,\textcolor{red}{-2.05}} & \rqonecell{6.16} & \rqtwocell{4.79}{\textcolor{green!50!black}{-1.37}} & \rqtwocell{6.85}{\textcolor{red}{+0.68};\,\textcolor{red}{+2.05}} \\
\addlinespace[2pt]
\rqonecell{Seed-Evolving}
& \rqonecell{49.49}
& \rqtwocell{52.53}{\textcolor{green!50!black}{+3.03}}
& \rqtwocell{45.45}{\textcolor{red}{-4.04};\,\textcolor{red}{-7.07}}
& \rqonecell{97.95}
& \rqtwocell{\textbf{97.95}}{0.00}
& \rqtwocell{\textbf{97.26}}{\textcolor{red}{-0.68};\,\textcolor{red}{-0.68}}
& \rqonecell{0.68}
& \rqtwocell{\textbf{1.37}}{\textcolor{red}{+0.68}}
& \rqtwocell{\textbf{1.37}}{\textcolor{red}{+0.68};\,0.00} \\
\addlinespace[2pt]
\rqonecell{Qwen3.8-27B}
& \rqonecell{\textbf{59.60}}
& \rqtwocell{\textbf{66.67}}{\textcolor{green!50!black}{+7.07}}
& \rqtwocell{\textbf{65.66}}{\textcolor{green!50!black}{+6.06};\,\textcolor{red}{-1.01}}
& \rqonecell{95.89}
& \rqtwocell{95.21}{\textcolor{red}{-0.68}}
& \rqtwocell{93.84}{\textcolor{red}{-2.05};\,\textcolor{red}{-1.37}}
& \rqonecell{2.74}
& \rqtwocell{4.11}{\textcolor{red}{+1.37}}
& \rqtwocell{4.79}{\textcolor{red}{+2.05};\,\textcolor{red}{+0.68}} \\
\addlinespace[2pt]
\rqonecell{Qwen3.7-Plus} & \rqonecell{57.58} & \rqtwocell{63.64}{\textcolor{green!50!black}{+6.06}} & \rqtwocell{64.65}{\textcolor{green!50!black}{+7.07};\,\textcolor{green!50!black}{+1.01}} & \rqonecell{96.58} & \rqtwocell{94.52}{\textcolor{red}{-2.05}} & \rqtwocell{95.21}{\textcolor{red}{-1.37};\,\textcolor{green!50!black}{+0.68}} & \rqonecell{2.05} & \rqtwocell{3.42}{\textcolor{red}{+1.37}} & \rqtwocell{3.42}{\textcolor{red}{+1.37};\,0.00} \\
\addlinespace[2pt]
\rqonecell{Qwen3.5-9B}
& \rqonecell{24.24}
& \rqtwocell{42.42}{\textcolor{green!50!black}{+18.18}}
& \rqtwocell{34.34}{\textcolor{green!50!black}{+10.10};\,\textcolor{red}{-8.08}}
& \rqonecell{\textbf{98.63}}
& \rqtwocell{90.41}{\textcolor{red}{-8.22}}
& \rqtwocell{91.10}{\textcolor{red}{-7.53};\,\textcolor{green!50!black}{+0.68}}
& \rqonecell{\textbf{0.00}}
& \rqtwocell{7.53}{\textcolor{red}{+7.53}}
& \rqtwocell{7.53}{\textcolor{red}{+7.53};\,0.00} \\
\addlinespace[2pt]
\rqonecell{DeepSeek-V4} & \rqonecell{36.36} & \rqtwocell{48.48}{\textcolor{green!50!black}{+12.12}} & \rqtwocell{46.46}{\textcolor{green!50!black}{+10.10};\,\textcolor{red}{-2.02}} & \rqonecell{95.89} & \rqtwocell{95.89}{0.00} & \rqtwocell{91.78}{\textcolor{red}{-4.11};\,\textcolor{red}{-4.11}} & \rqonecell{2.05} & \rqtwocell{2.05}{0.00} & \rqtwocell{5.48}{\textcolor{red}{+3.42};\,\textcolor{red}{+3.42}} \\
\addlinespace[2pt]
\rqonecell{GUI-OWL-7B} & \rqonecell{26.26} & \rqtwocell{37.37}{\textcolor{green!50!black}{+11.11}} & \rqtwocell{43.43}{\textcolor{green!50!black}{+17.17};\,\textcolor{green!50!black}{+6.06}} & \rqonecell{91.10} & \rqtwocell{87.67}{\textcolor{red}{-3.42}} & \rqtwocell{87.67}{\textcolor{red}{-3.42};\,0.00} & \rqonecell{0.68} & \rqtwocell{4.11}{\textcolor{red}{+3.42}} & \rqtwocell{2.74}{\textcolor{red}{+2.05};\,\textcolor{green!50!black}{-1.37}} \\
\midrule
\rqonecell{Average} & \rqonecell{46.02} & \rqtwocell{53.87}{\textcolor{green!50!black}{+7.86}} & \rqtwocell{52.64}{\textcolor{green!50!black}{+6.62};\,\textcolor{red}{-1.23}} & \rqonecell{95.21} & \rqtwocell{93.53}{\textcolor{red}{-1.67}} & \rqtwocell{92.39}{\textcolor{red}{-2.82};\,\textcolor{red}{-1.14}} & \rqonecell{2.05} & \rqtwocell{3.88}{\textcolor{red}{+1.83}} & \rqtwocell{4.34}{\textcolor{red}{+2.28};\,\textcolor{red}{+0.46}} \\
\bottomrule
\end{tabular*}
\begin{tablenotes}[flushleft]
\scriptsize
\item[1] DeepSeek-V4 is an abbreviation for DeepSeek-V4-Flash-Vision.
\end{tablenotes}
\end{threeparttable}
\end{table*}
\endgroup

\paragraph{Effect on swipe decisions.}
As shown in Table~\ref{tab:rq2_swipe_planning}, adding \MIG improves swipe prediction recall for all nine VLMs.
On average, swipe prediction recall increases from 46.02\% under \textit{Vanilla} to 53.87\% under \textit{+\MIG}, with gains ranging from 2.02 to 18.18 percentage points.
The largest improvement is observed on Qwen3.5-9B (+18.18 points), followed by DeepSeek-V4-Flash-Vision (+12.12) and GUI-OWL-7B (+11.11).
These results show that explicit scrollable regions consistently help VLMs recognize when a swipe is required, including models with substantially different vanilla performance.

\paragraph{Effect on click decisions.}
The improvement in swipe decisions introduces only limited interference with click decisions.
While \MIG increases average swipe recall by 7.85 percentage points, click recall decreases by only 1.68 points, from 95.21\% to 93.53\%.
Meanwhile, the Click$\rightarrow$Swipe rate increases modestly from 2.05\% to 3.88\% (+1.83 points).
These results show that \MIG makes agents substantially more likely to swipe when exploration is needed, without causing them to swipe unnecessarily.

\paragraph{Comparison with ground-truth regions.}
Replacing \MIG predictions with ground-truth scrollable regions brings only limited additional benefit to swipe planning.
On average, Swipe Recall under \textit{+Ground Truth} is 52.64\%, close to the 53.87\% achieved by \textit{+\MIG}.
This small gap suggests that \MIG's predicted regions already capture most of the scrollability information that is useful for swipe decisions. At the same time, the limited additional benefit from ground-truth regions indicates that scrollable-region perception is not the only factor affecting swipe planning. Even with accurate region cues, agents still need to determine how to incorporate them into task-level action selection.

\subsection{RQ3: Plug-in Effectiveness, Overhead, and Side Effects}

\textbf{RQ3} evaluates how \MIG affects the end-to-end task automation performance of GUI agents.
This experiment examines whether scrollable-region perception can lead to higher end-to-end task success and assesses the additional execution cost.
The three VLMs with the highest swipe prediction recall in \textbf{RQ2}'s \textit{+\MIG} setting are selected for end-to-end evaluation.
GUI agents powered by these VLMs are then evaluated under \textit{Vanilla} and \textit{+\MIG} on the 18 swipe-dependent tasks, with each task run three times per agent and setting and success assessed manually for each run.
Table~\ref{tab:rq3_agent_integration} reports task success rate (SR) and its standard deviation across rounds ($\sigma$(SR)), average interaction steps for all runs and for successful runs, and per-run averages of execution time and input/output token usage.

\begingroup
\newcommand{\rqthreecell}[2]{#1\,{\scriptsize\ensuremath{(#2)}}}
\begin{table*}[t]
\centering
\footnotesize
\caption{Online task performance and execution cost with and without \MIG.}
\label{tab:rq3_agent_integration}
\begin{threeparttable}
\setlength{\tabcolsep}{1.75pt}
\renewcommand{\arraystretch}{1.05}
\begin{tabular*}{\linewidth}{@{\extracolsep{\fill}}cccccccc}
\toprule
\makecell[c]{\textbf{Agent}} & \makecell[c]{\textbf{Mode}} & \makecell[c]{\textbf{SR} (\%)} & \makecell[c]{$\boldsymbol{\sigma}$(\textbf{SR})} & \makecell[c]{\textbf{Total/Succ.}\\(steps/task)} & \makecell[c]{\textbf{Time}\\(s/task)} & \makecell[c]{\textbf{Input Token}\\(/task)} & \makecell[c]{\textbf{Output Token}\\(/task)} \\
\midrule
\multirow{2}{*}{Qwen3.8-27B}
 & V & 70.37 & 5.24 & 10.93 / 10.84 & 85.84 & 75,823 & 822 \\
 & H & \rqthreecell{88.89}{\textcolor{green!50!black}{+18.52}} & 4.54 & 10.02 / 9.29 & 129.77 & 69,472 + 2,261 & 757 + 531 \\
\midrule
\multirow{2}{*}{Qwen3.7-Plus}
 & V & 74.07 & 2.62 & 11.06 / 9.53 & 88.79 & 76,800 & 909 \\
 & H & \rqthreecell{88.89}{\textcolor{green!50!black}{+14.81}} & 4.54 & 10.07 / 9.50 & 122.19 & 69,847 + 1,805 & 816 + 418 \\
\midrule
\multirow{2}{*}{GPT-6-Luna}
 & V & 50.00 & 12.00 & 10.11 / 9.52 & 97.41 & 112,739 & 1,058 \\
 & H & \rqthreecell{70.37}{\textcolor{green!50!black}{+20.37}} & 6.93 & 11.59 / 10.79 & 168.48 & 128,915 + 2,517 & 1,119 + 598 \\
\bottomrule
\end{tabular*}
\begin{tablenotes}[flushleft]
\scriptsize
\item[1] Total/Succ. reports mean steps over all tasks / successful tasks.
\item[2] $\sigma$(SR) is the population standard deviation of three round-level SRs. Parenthesized SR gains for \textit{+\MIG} over \textit{Vanilla} use original success counts.
\item[3] Token usage is reported separately for the agent and \MIG, shown as <agent $+$ \MIG>: agent tokens cover decision-making and memory generation, while \MIG tokens correspond to region-annotation inference. 
\end{tablenotes}
\end{threeparttable}
\end{table*}
\endgroup

\paragraph{Task completion and consistency.}
All three agents achieve higher task success rates under \textit{+\MIG} on swipe-dependent tasks, with the mean rising from 64.81\% to 82.72\% (+17.90 percentage points).
This agreement with \textbf{RQ2} suggests that region cues remain useful when swipe decisions are embedded in complete task trajectories.
GPT-6-Luna shows the largest gain in task success rate, rising from 50.00\% to 70.37\%, while both Qwen agents reach 88.89\%, showing that scrollable-region perception benefits even agents with stronger baseline performance.

Moreover, \MIG reduces the standard deviation of task success rates across evaluation rounds for Qwen3.8-27B and GPT-6-Luna, particularly the latter with its lower baseline success rate, indicating more consistent task completion across repeated runs.
Nevertheless, we acknowledge that \MIG's prediction errors can contribute to task failures. A representative case is discussed in Section~\ref{sec:rq3_case_study}.

\paragraph{Interaction efficiency.}
Given GUI agents' ability to recover from suboptimal actions during task execution, we evaluate whether the proactive exploration encouraged by \MIG reduces the number of interactions required to complete a task.
Overall, \textit{+\MIG} reduces the mean number of interactions per task, with the effect varying across agents: Qwen3.8-27B and Qwen3.7-Plus complete more tasks with fewer interactions, whereas GPT-6-Luna follows longer trajectories.

\paragraph{Execution cost.}
Adding \MIG naturally introduces additional execution overhead, increasing active execution time by 49.46 seconds per run on average, with increases ranging from 33.40 to 71.07 seconds across agents. This overhead increases the time required to obtain a task outcome and may reduce the number of attempts possible under a fixed time budget. These measurements already account for cache reuse, with hit rates ranging from 62.30\% to 70.34\% among calls carrying cache metadata.

However, \MIG does not increase token consumption for the Qwen-based agents. Their lower token usage, including region-annotation inference, is consistent with the shorter interaction trajectories enabled by more targeted exploration, suggesting that improved exploration efficiency can also reduce the token cost of agent decision-making and  reflection. Overall, while \MIG improves task success at the cost of additional execution time, its more accurate exploration can reduce token consumption for capable agents.

\section{Discussion}
\label{sec:discussion}

This section first uses a case study to show that, although scrollable-region perception has proven valuable for GUI agents, integrating it through supplementary cues has inevitable limitations.
The case illustrates how \MIG can degrade an agent's task performance.
It then considers threats to validity arising from manual annotation and variations in agent performance.

\subsection{Case Study: Towards Integrated Perception and Planning}
\label{sec:rq3_case_study}

In RQ3, we observe that although \MIG is designed to preserve agents' decision-making autonomy, its predictions can still misguide their exploration. To better understand this phenomenon, we analyze a failed execution run on TrainPal's homepage.
The task requires locating \textit{KORAIL PASS} through the \textit{Global Pass} section on TrainPal's homepage.
In a \textbf{+\MIG} run, the Qwen3.8-27B-based agent fails to locate \textit{KORAIL PASS}, illustrating how incomplete region cues can interfere with its exploration.

\begin{figure}[t]
    \centering
    \subfigure[Vanilla exploration]{%
        \includegraphics[height=0.485\columnwidth]{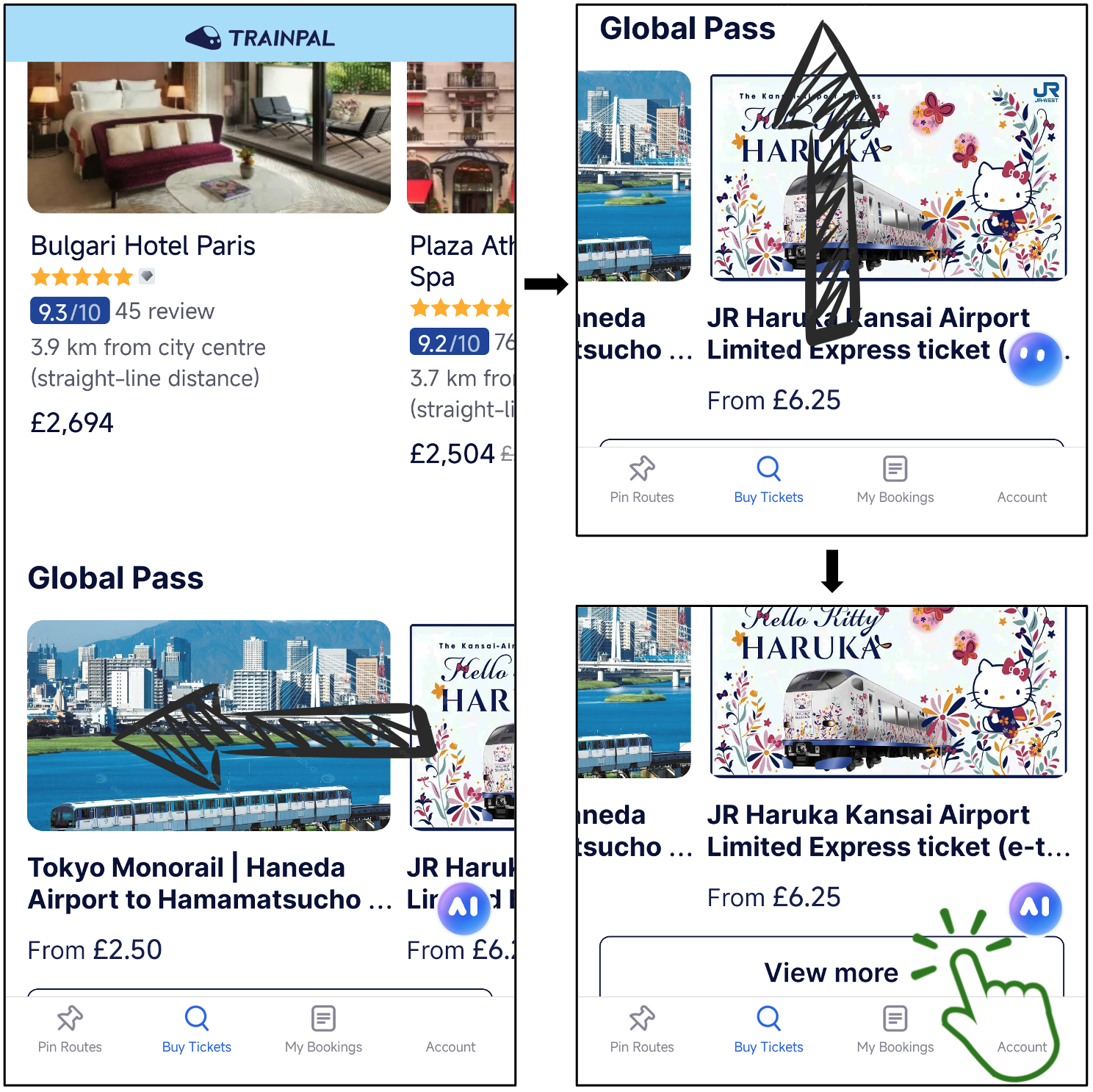}%
        \label{fig:trainpal_vanilla}%
    }\hfill
    \subfigure[Exploration with \MIG]{%
        \includegraphics[height=0.485\columnwidth]{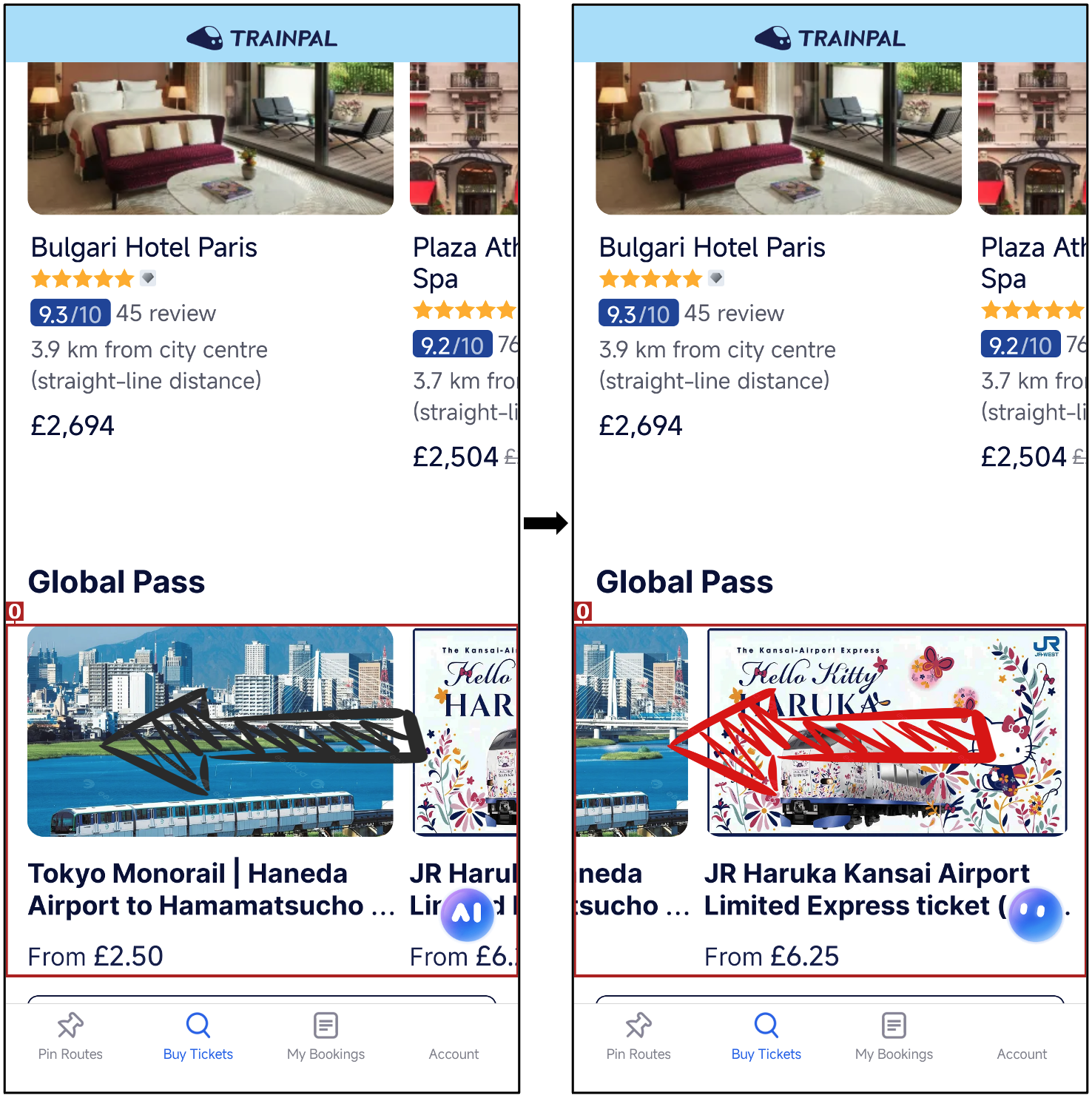}%
        \label{fig:trainpal_guided}%
    }
    \caption{Comparison of Qwen3.8-27B's exploration in TrainPal: (a) Vanilla exploration reveals \textit{View more} through page scrolling, whereas (b) exploration with \MIG remains confined to the carousel. Arrows indicate swipe actions, and red boxes mark scrollable regions predicted by \MIG.}
    \label{fig:rq3_trainpal_failure}
\end{figure}

\begingroup
\setlength{\emergencystretch}{1em} Figure~\ref{fig:trainpal_vanilla} shows the carousel endpoints and the route to the full catalogue for Vanilla exploration.
Neither end of the carousel displays \textit{KORAIL PASS}.
In a successful Vanilla run, the agent swipes up on the page to reveal \textit{View more}, which is initially concealed beneath the bottom navigation bar.
Selecting \textit{View more} opens the full \textit{Global Pass} catalogue, which lists the available passes and tickets by category.
Within the catalogue, the agent swipes left across the category bar, selects \textit{South Korea}, and opens \textit{KORAIL PASS}.
\par\endgroup

In the illustrated steps of the failed run, \MIG highlights the horizontally scrollable preview while omitting the surrounding vertically scrollable page.
The \textit{View more} label remains concealed beneath the bottom navigation bar, therefore reaching the full catalogue requires further vertical page scrolling. 
As shown in Figure~\ref{fig:trainpal_guided}, the agent swipes left from the highlighted carousel's left-end view to its right-end view.
This transition changes the carousel's visible content, but further leftward swipes at the right boundary produce no visible change and do not reveal \textit{KORAIL PASS}.
The agent continues exploring the carousel instead of scrolling the page to expose \textit{View more}, and ultimately returns \texttt{finish} on the homepage without locating the requested pass.

This case indicates that preserving an agent's decision-making autonomy does not fully prevent incomplete perception from misleading its planning.
Although \MIG only provides supplementary guidance without enforcing any action, the agent may still rely too heavily on these cues and fail to revise its search strategy after repeated unsuccessful interactions.
This suggests future research on internalizing scrollable-region perception into GUI agents through large-scale joint training with task planning, so that agents can systematically learn when to explore, where to explore, and when to stop exploring.

\subsection{Threats to Validity}
\label{sec:threats}

\paragraph{Annotation subjectivity.}
Since scrollable regions may lack clear visual boundaries, manual annotations are susceptible to omissions and errors.
Local or nested regions can be overlooked, and annotators may disagree on their boundaries, affecting both training supervision and the reference annotations used in RQ1.
Although XML-derived candidates provide an initial set of regions for inspection, the candidate set may omit scrollable regions or include non-scrollable ones.
To mitigate this threat, two authors independently inspect each screenshot, remove invalid candidates, and add missed regions.
A third author resolves disagreements to finalize the annotations.
Although some annotation noise may remain, candidate-assisted inspection helps identify potential regions, while independent checks and disagreement resolution reduce reliance on individual judgments.

\paragraph{Stability of effectiveness.}
Changes in UI states and variations in agent behavior may affect the observed performance gains from \MIG.
The same task may encounter different interface states or action sequences across executions, so a single run may overestimate or underestimate the benefit of the plug-in.
We acknowledge this threat and seek to mitigate it by evaluating diverse test tasks with multiple agents.
This broader evaluation reduces dependence on a particular task or agent's decision patterns.
In RQ3, \MIG improves the task success rate of all three evaluated agents.
Moreover, each task in RQ3 is executed three times, and we report the mean within-agent population standard deviation of round-level task success rates (5.33 percentage points under \textbf{+\MIG}, obtained by averaging the three agents' unrounded standard deviations over the three evaluation rounds; Table~\ref{tab:rq3_agent_integration}).
These repeated runs allow us to assess whether the observed gains persist across executions and quantify the remaining variation.

\section{Related Work}
\label{sec:related}

\paragraph{Interactive region perception.}

Identifying and localizing interactive regions is fundamental to GUI task automation.
Structured approaches use UI hierarchies and element attributes to expose interaction targets~\cite{wen2024autodroid,deng2023mind2web}, while visual approaches ground actions directly in screenshots.
ScreenAI~\cite{baechler2024screenai} studies screen understanding and UI element annotation, and SeeClick~\cite{cheng2024seeclick} develops GUI grounding through visual pretraining.
Building on visual localization, ShowUI~\cite{lin2024showui}, OS-Atlas~\cite{wu2025osatlas}, and UIPro~\cite{li2025uipro} further connect grounding with GUI action prediction.
ScreenSpot-Pro~\cite{li2025screenspotpro} evaluates these localization capabilities in high-resolution professional interfaces.

Beyond locating interaction targets, existing work also investigates what interactions UI regions support.
Schoop et al.~\cite{schoop2022tappability} predict whether users perceive UI elements as tappable, while Ferret-UI~\cite{you2024ferretui} supports UI function inference and goal-oriented interaction conversations.
OmniParser~\cite{lu2024omniparser} combines interactable-element detection with functional descriptions to support action grounding.
Complementing such perception methods, Set-of-Mark prompting~\cite{yang2023som} overlays marks on image regions so that a multimodal model can refer to them explicitly.
These methods improve the localization, functional understanding, and explicit referencing of visible interaction targets.
However, none specifically focuses on identifying scrollable containers that can reveal content beyond the current viewport.

\paragraph{GUI exploration and planning.}

Existing work guides GUI exploration to improve testing efficiency.
Stoat~\cite{su2017stoat} uses heuristic exploration strategies and GUI models to guide test generation.
LLMDroid~\cite{wang2025llmdroid} uses general-purpose LLMs to direct existing testing tools toward additional functionality when coverage growth slows.
Guardian~\cite{ran2024guardian} uses runtime action constraints and replanning to guide exploration toward UI states relevant to testing objectives.
These approaches focus on selecting and adjusting exploration actions to improve testing efficiency, rather than explicitly learning the interaction semantics that determine which UI regions support scrolling.

Recent GUI agents also improve planning through interaction history, reflection, and proactive exploration~\cite{liu2025guirise,wu2025guireflection,li2025mobileuse}.
In particular, MobileUse~\cite{li2025mobileuse} combines hierarchical reflection with self-planned exploration to build knowledge of unfamiliar environments.
These mechanisms address how agents accumulate and use experience to guide subsequent actions.
However, for complex apps, building such knowledge can require extensive preliminary exploration without guaranteeing useful guidance for a given task.
\MIG instead provides scrollable-region cues directly from the current screenshot and is shown to be effective with modest overhead.

\section{Conclusion}\label{sec:conclusion}
Motivated by the huge gap between agents' GUI and CLI task performance, this paper identifies invisible content as a key bottleneck for GUI tasks.
When the correct target is hidden, agents often click an incorrect visible target instead of exploring.
Built on a manually annotated dataset, \MIG provides non-intrusive scrollable-region cues that support hidden-area exploration while preserving GUI agents' decision-making autonomy.
Experiments demonstrate that such scrollable-region prompts improve swipe selection accuracy by 17.1\% and result in a 27.6\% increase in the end-to-end task completion rate.
These findings suggest that making exploration opportunities explicit is a practical direction for improving GUI agent harnesses under partial observability.

\section{DATA AVAILABILITY}
All of the source code, dataset, and experimental records for \MIG are available at \url{https://figshare.com/s/c39e36c1c8fb0b48a1b6}.

\bibliographystyle{ACM-Reference-Format}
\bibliography{ui_testing}

@preamble{"\providecommand{\natexlab}[1]{}\renewcommand{\natexlab}[1]{}"}

@book{goodfellow2016deep,
  author       = {Ian Goodfellow and Yoshua Bengio and Aaron Courville},
  title        = {Deep Learning},
  publisher    = {MIT Press},
  year         = {2016}
}

@online{googleplayranking,
  title        = {App Discovery and Ranking},
  year         = {Accessed: 2026},
  url          = {https://support.google.com/googleplay/android-developer/answer/9958766}
}

@online{androidxmlhierarchy,
  title        = {{UI Automator}: {AccessibilityNodeInfoDumper}},
  year         = {Accessed: 2026},
  url          = {https://android.googlesource.com/platform/frameworks/testing/+/master/uiautomator/library/core-src/com/android/uiautomator/core/AccessibilityNodeInfoDumper.java}
}

@online{materialdesignlists,
  title        = {Lists---Material Design},
  year         = {Accessed: 2026},
  url          = {https://m1.material.io/components/lists.html}
}

@online{androidadb,
  title        = {Android Debug Bridge (adb)},
  year         = {Accessed: 2026},
  url          = {https://developer.android.com/tools/adb}
}

@inproceedings{reinecke2013aesthetics,
  author       = {Katharina Reinecke and Tom Yeh and Luke Miratrix and Rahmatri Mardiko and
                  Yuechen Zhao and Jenny Liu and Krzysztof Z. Gajos},
  title        = {Predicting Users' First Impressions of Website Aesthetics with a Quantification of Perceived Visual Complexity and Colorfulness},
  booktitle    = {Proceedings of the SIGCHI Conference on Human Factors in Computing Systems},
  series       = {CHI '13},
  pages        = {2049--2058},
  publisher    = {Association for Computing Machinery},
  year         = {2013},
}

@inproceedings{pfeuffer2018grid,
  author       = {Ken Pfeuffer and Yang Li},
  title        = {Analysis and Modeling of Grid Performance on Touchscreen Mobile Devices},
  booktitle    = {Proceedings of the 2018 CHI Conference on Human Factors in Computing Systems},
  series       = {CHI '18},
  articleno    = {288},
  numpages     = {12},
  publisher    = {Association for Computing Machinery},
  year         = {2018},
}

@inproceedings{yang2023intercode,
  author       = {John Yang and Akshara Prabhakar and Karthik Narasimhan and Shunyu Yao},
  title        = {{InterCode}: Standardizing and Benchmarking Interactive Coding with Execution Feedback},
  booktitle    = {Advances in Neural Information Processing Systems},
  volume       = {36},
  pages        = {23826--23854},
  publisher    = {Curran Associates, Inc.},
  year         = {2023},
}

@inproceedings{yang2024sweagent,
  author       = {John Yang and Carlos Jimenez and Alexander Wettig and Kilian Lieret and
                  Shunyu Yao and Karthik Narasimhan and Ofir Press},
  title        = {{SWE-agent}: Agent-Computer Interfaces Enable Automated Software Engineering},
  booktitle    = {Advances in Neural Information Processing Systems},
  volume       = {37},
  pages        = {50528--50652},
  publisher    = {Curran Associates, Inc.},
  year         = {2024},
}

@inproceedings{zhang2023appagent,
  author       = {Chi Zhang and Zhao Yang and Jiaxuan Liu and Yanda Li and Yucheng Han and Xin Chen and Zebiao Huang and Bin Fu and Gang Yu},
  title        = {{AppAgent}: Multimodal Agents as Smartphone Users},
  booktitle    = {Proceedings of the 2025 CHI Conference on Human Factors in Computing Systems},
  year         = {2025},
  pages        = {1--20},
}

@article{wang2024mobileagent,
  author       = {Junyang Wang and
                  Haiyang Xu and
                  Jiabo Ye and
                  Ming Yan and
                  Weizhou Shen and
                  Ji Zhang and
                  Fei Huang and
                  Jitao Sang},
  title        = {Mobile-Agent: Autonomous Multi-Modal Mobile Device Agent with Visual Perception},
  journal      = {CoRR},
  volume       = {abs/2401.16158},
  year         = {2024},
  note         = {Presented at the ICLR 2024 Workshop on Large Language Model (LLM) Agents}
}

@inproceedings{cheng2024seeclick,
  author       = {Kanzhi Cheng and
                  Qiushi Sun and
                  Yougang Chu and
                  Fangzhi Xu and
                  Yantao Li and
                  Jianbing Zhang and
                  Zhiyong Wu},
  title        = {{SeeClick}: Harnessing {GUI} Grounding for Advanced Visual {GUI} Agents},
  booktitle    = {Proceedings of the 62nd Annual Meeting of the Association for Computational Linguistics (Volume 1: Long Papers)},
  pages        = {9313--9332},
  publisher    = {Association for Computational Linguistics},
  year         = {2024},
}

@inproceedings{rawles2025androidworld,
  author       = {Christopher Rawles and
                  Sarah Clinckemaillie and
                  Yifan Chang and
                  Jonathan Waltz and
                  Gabrielle Lau and
                  Marybeth Fair and
                  Alice Li and
                  William Bishop and
                  Wei Li and
                  Folawiyo Campbell{-}Ajala and
                  Daniel Toyama and
                  Robert Berry and
                  Divya Tyamagundlu and
                  Timothy Lillicrap and
                  Oriana Riva},
  title        = {{AndroidWorld}: A Dynamic Benchmarking Environment for Autonomous Agents},
  booktitle    = {The Thirteenth International Conference on Learning Representations},
  year         = {2025},
}

@inproceedings{kong2026mobileworld,
  author       = {Quyu Kong and Xu Zhang and Zhenyu Yang and Nolan Gao and
                  Chen Liu and Panrong Tong and Chenglin Cai and Hanzhang Zhou and
                  Jianan Zhang and Liangyu Chen and Zhidan Liu and Steven Hoi and Yue Wang},
  title        = {{MobileWorld}: Benchmarking Autonomous Mobile Agents in Agent-User Interactive and {MCP}-Augmented Environments},
  booktitle    = {Proceedings of the 64th Annual Meeting of the Association for Computational Linguistics (Volume 1: Long Papers)},
  pages        = {6142--6167},
  publisher    = {Association for Computational Linguistics},
  year         = {2026},
}

@inproceedings{xie2024osworld,
  author       = {Tianbao Xie and
                  Danyang Zhang and
                  Jixuan Chen and
                  Xiaochuan Li and
                  Siheng Zhao and
                  Ruisheng Cao and
                  Toh Jing Hua and
                  Zhoujun Cheng and
                  Dongchan Shin and
                  Fangyu Lei and
                  Yitao Liu and
                  Yiheng Xu and
                  Shuyan Zhou and
                  Silvio Savarese and
                  Caiming Xiong and
                  Victor Zhong and
                  Tao Yu},
  title        = {{OSWorld}: Benchmarking Multimodal Agents for Open-Ended Tasks in Real Computer Environments},
  booktitle    = {Advances in Neural Information Processing Systems},
  year         = {2024},
}

@article{qin2025uitars,
  author       = {Yujia Qin and
                  Yining Ye and
                  Junjie Fang and
                  Haoming Wang and
                  Shihao Liang and
                  Shizuo Tian and
                  Junda Zhang and
                  Jiahao Li and
                  Yunxin Li and
                  Shijue Huang and
                  Wanjun Zhong and
                  Kuanye Li and
                  Jiale Yang and
                  Yu Miao and
                  Woyu Lin and
                  Longxiang Liu and
                  Xu Jiang and
                  Qianli Ma and
                  Jingyu Li and
                  Xiaojun Xiao and
                  Kai Cai and
                  Chuang Li and
                  Yaowei Zheng and
                  Chaolin Jin and
                  Chen Li and
                  Xiao Zhou and
                  Minchao Wang and
                  Haoli Chen and
                  Zhaojian Li and
                  Haihua Yang and
                  Haifeng Liu and
                  Feng Lin and
                  Tao Peng and
                  Xin Liu and
                  Guang Shi},
  title        = {{UI-TARS}: Pioneering Automated {GUI} Interaction with Native Agents},
  journal      = {CoRR},
  volume       = {abs/2501.12326},
  year         = {2025},
}

@inproceedings{deng2023mind2web,
  author       = {Xiang Deng and
                  Yu Gu and
                  Boyuan Zheng and
                  Shijie Chen and
                  Samuel Stevens and
                  Boshi Wang and
                  Huan Sun and
                  Yu Su},
  title        = {{Mind2Web}: Towards a Generalist Agent for the Web},
  booktitle    = {Advances in Neural Information Processing Systems},
  year         = {2023},
}

@inproceedings{lu2025guiodyssey,
  author       = {Quanfeng Lu and Wenqi Shao and Zitao Liu and Lingxiao Du and
                  Fanqing Meng and Boxuan Li and Botong Chen and Siyuan Huang and
                  Kaipeng Zhang and Ping Luo},
  title        = {{GUIOdyssey}: A Comprehensive Dataset for Cross-App {GUI} Navigation on Mobile Devices},
  booktitle    = {Proceedings of the IEEE/CVF International Conference on Computer Vision},
  pages        = {22404--22414},
  year         = {2025},
}

@inproceedings{rawles2023aitw,
  author       = {Christopher Rawles and
                  Alice Li and
                  Daniel Rodriguez and
                  Oriana Riva and
                  Timothy P. Lillicrap},
  title        = {{AndroidInTheWild}: A Large-Scale Dataset for Android Device Control},
  booktitle    = {Advances in Neural Information Processing Systems},
  year         = {2023},
}

@inproceedings{shaw2023pixels,
  author       = {Peter Shaw and
                  Mandar Joshi and
                  James Cohan and
                  Jonathan Berant and
                  Panupong Pasupat and
                  Hexiang Hu and
                  Urvashi Khandelwal and
                  Kenton Lee and
                  Kristina Toutanova},
  title        = {From Pixels to {UI} Actions: Learning to Follow Instructions via Graphical User Interfaces},
  year         = {2023},
  booktitle    = {Advances in Neural Information Processing Systems},
  volume       = {36},
  pages        = {34354--34370},
  publisher    = {Curran Associates, Inc.},
}

@inproceedings{wu2025osatlas,
  author       = {Zhiyong Wu and
                  Zhenyu Wu and
                  Fangzhi Xu and
                  Yian Wang and
                  Qiushi Sun and
                  Chengyou Jia and
                  Kanzhi Cheng and
                  Zichen Ding and
                  Liheng Chen and
                  Paul Pu Liang and
                  Yu Qiao},
  title        = {{OS-ATLAS}: Foundation Action Model for Generalist {GUI} Agents},
  booktitle    = {The Thirteenth International Conference on Learning Representations},
  year         = {2025},
}

@inproceedings{you2024ferretui,
  author       = {Keen You and
                  Haotian Zhang and
                  Eldon Schoop and
                  Floris Weers and
                  Amanda Swearngin and
                  Jeffrey Nichols and
                  Yinfei Yang and
                  Zhe Gan},
  title        = {Ferret-UI: Grounded Mobile {UI} Understanding with Multimodal {LLMs}},
  year         = {2024},
  booktitle    = {Computer Vision -- ECCV 2024},
  pages        = {240--255},
  publisher    = {Springer Nature Switzerland},
}

@inproceedings{baechler2024screenai,
  author       = {Gilles Baechler and
                  Srinivas Sunkara and
                  Maria Wang and
                  Fedir Zubach and
                  Hassan Mansoor and
                  Vincent Etter and
                  Victor Carbune and
                  Jason Lin and
                  Jindong Chen and
                  Abhanshu Sharma},
  title        = {{ScreenAI}: A Vision-Language Model for {UI} and Infographics Understanding},
  booktitle    = {Proceedings of the Thirty-Third International Joint Conference on Artificial Intelligence},
  pages        = {3058--3068},
  publisher    = {ijcai.org},
  year         = {2024},
}

@inproceedings{lin2024showui,
  author       = {Kevin Qinghong Lin and Linjie Li and Difei Gao and Zhengyuan Yang and Shiwei Wu and Zechen Bai and Stan Weixian Lei and Lijuan Wang and Mike Zheng Shou},
  title        = {{ShowUI}: One Vision-Language-Action Model for {GUI} Visual Agent},
  booktitle    = {Proceedings of the IEEE/CVF Conference on Computer Vision and Pattern Recognition},
  year         = {2025},
  pages        = {19498--19508},
}

@inproceedings{xu2024androidlab,
  author       = {Yifan Xu and
                  Xiao Liu and
                  Xueqiao Sun and
                  Siyi Cheng and
                  Hao Yu and
                  Hanyu Lai and
                  Shudan Zhang and
                  Dan Zhang and
                  Jie Tang and
                  Yuxiao Dong},
  title        = {{AndroidLab}: Training and Systematic Benchmarking of Android Autonomous Agents},
  year         = {2025},
  booktitle    = {Proceedings of the 63rd Annual Meeting of the Association for Computational Linguistics (Volume 1: Long Papers)},
  pages        = {2144--2166},
  publisher    = {Association for Computational Linguistics},
}

@article{ye2025mobileagentv3,
  author       = {Jiabo Ye and Xi Zhang and Haiyang Xu and Haowei Liu and Junyang Wang and
                  Zhaoqing Zhu and Ziwei Zheng and Feiyu Gao and Junjie Cao and Zhengxi Lu and
                  Jitong Liao and Qi Zheng and Fei Huang and Jingren Zhou and Ming Yan},
  title        = {Mobile-Agent-v3: Fundamental Agents for {GUI} Automation},
  journal      = {CoRR},
  volume       = {abs/2508.15144},
  year         = {2025},
}

@inproceedings{wu2025guiactor,
  author       = {Qianhui Wu and
                  Kanzhi Cheng and
                  Rui Yang and
                  Chaoyun Zhang and
                  Jianwei Yang and
                  Huiqiang Jiang and
                  Jian Mu and
                  Baolin Peng and
                  Bo Qiao and
                  Reuben Tan and
                  Si Qin and
                  Lars Liden and
                  Qingwei Lin and
                  Huan Zhang and
                  Tong Zhang and
                  Jianbing Zhang and
                  Dongmei Zhang and
                  Jianfeng Gao},
  title        = {{GUI-Actor}: Coordinate-Free Visual Grounding for {GUI} Agents},
  booktitle    = {Advances in Neural Information Processing Systems},
  year         = {2025},
}

@inproceedings{wu2025guireflection,
  author       = {Penghao Wu and
                  Shengnan Ma and
                  Bo Wang and
                  Jiaheng Yu and
                  Lewei Lu and
                  Ziwei Liu},
  title        = {{GUI-Reflection}: Empowering Multimodal {GUI} Models with Self-Reflection Behavior},
  booktitle    = {Advances in Neural Information Processing Systems},
  year         = {2025},
}

@article{li2025mobileuse,
  author       = {Li, Ning and
                  Qu, Xiangmou and
                  Zhou, Jiamu and
                  Wen, Muning and
                  Du, Kounianhua and
                  Lou, Xingyu and
                  Peng, Qiuying and
                  Wang, Jun and
                  Zhang, Weinan},
  title        = {{MobileUse}: A Hierarchical Reflection-Driven {GUI} Agent for Autonomous Mobile Operation},
  journal      = {Advances in Neural Information Processing Systems},
  volume       = {38},
  pages        = {40361--40388},
  year         = {2026},
}

@article{liu2025guirise,
  author       = {Liu, Tao and
                  Wang, Chongyu and
                  Li, Rongjie and
                  Yu, Yingchen and
                  He, Xuming and
                  Bai, Song},
  title        = {Gui-rise: Structured reasoning and history summarization for gui navigation},
  journal      = {Advances in Neural Information Processing Systems},
  volume       = {38},
  pages        = {165132--165156},
  year         = {2026},
}

@inproceedings{yao2023react,
  author       = {Shunyu Yao and
                  Jeffrey Zhao and
                  Dian Yu and
                  Nan Du and
                  Izhak Shafran and
                  Karthik Narasimhan and
                  Yuan Cao},
  title        = {{ReAct}: Synergizing Reasoning and Acting in Language Models},
  booktitle    = {The Eleventh International Conference on Learning Representations},
  year         = {2023},
}

@inproceedings{li2025screenspotpro,
  author       = {Kaixin Li and
                  Ziyang Meng and
                  Hongzhan Lin and
                  Ziyang Luo and
                  Yuchen Tian and
                  Jing Ma and
                  Zhiyong Huang and
                  Tat{-}Seng Chua},
  title        = {{ScreenSpot-Pro}: {GUI} Grounding for Professional High-Resolution Computer Use},
  year         = {2025},
  booktitle    = {Proceedings of the 33rd ACM International Conference on Multimedia},
  pages        = {8778--8786},
  publisher    = {Association for Computing Machinery},
}

@article{zhou2025maiuitechnicalreportrealworld,
  author       = {Hanzhang Zhou and Xu Zhang and Panrong Tong and Jianan Zhang and Liangyu Chen and Quyu Kong and Chenglin Cai and Chen Liu and Yue Wang and Jingren Zhou and Steven Hoi},
  title        = {MAI-UI Technical Report: Real-World Centric Foundation GUI Agents},
  journal      = {CoRR},
  volume       = {abs/2512.22047},
  year         = {2025}
}

@article{bai2025qwen3vltechnicalreport,
  author       = {Shuai Bai and Yuxuan Cai and Ruizhe Chen and Keqin Chen and Xionghui Chen and Zesen Cheng and Lianghao Deng and Wei Ding and Chang Gao and Chunjiang Ge and Wenbin Ge and Zhifang Guo and Qidong Huang and Jie Huang and Fei Huang and Binyuan Hui and Shutong Jiang and Zhaohai Li and Mingsheng Li and Mei Li and Kaixin Li and Zicheng Lin and Junyang Lin and Xuejing Liu and Jiawei Liu and Chenglong Liu and Yang Liu and Dayiheng Liu and Shixuan Liu and Dunjie Lu and Ruilin Luo and Chenxu Lv and Rui Men and Lingchen Meng and Xuancheng Ren and Xingzhang Ren and Sibo Song and Yuchong Sun and Jun Tang and Jianhong Tu and Jianqiang Wan and Peng Wang and Pengfei Wang and Qiuyue Wang and Yuxuan Wang and Tianbao Xie and Yiheng Xu and Haiyang Xu and Jin Xu and Zhibo Yang and Mingkun Yang and Jianxin Yang and An Yang and Bowen Yu and Fei Zhang and Hang Zhang and Xi Zhang and Bo Zheng and Humen Zhong and Jingren Zhou and Fan Zhou and Jing Zhou and Yuanzhi Zhu and Ke Zhu},
  title        = {Qwen3-VL Technical Report},
  journal      = {CoRR},
  volume       = {abs/2511.21631},
  year         = {2025}
}

@inproceedings{tang2025guig2gaussianrewardmodeling,
  title        = {GUI-G$^2$: Gaussian Reward Modeling for GUI Grounding},
  author       = {Fei Tang and Zhangxuan Gu and Zhengxi Lu and Xuyang Liu and Shuheng Shen and Changhua Meng and Wen Wang and Wenqi Zhang and Yongliang Shen and Weiming Lu and Jun Xiao and Yueting Zhuang},
  year         = {2026},
  booktitle    = {Proceedings of the AAAI Conference on Artificial Intelligence},
  volume       = {40},
  number       = {39},
  pages        = {33214--33222},
}

@article{venusteam2026uivenus15technicalreport,
  author       = {Venus Team and Changlong Gao and Zhangxuan Gu and Yulin Liu and Xinyu Qiu and Shuheng Shen and Yue Wen and Tianyu Xia and Zhenyu Xu and Zhengwen Zeng and Beitong Zhou and Xingran Zhou and Weizhi Chen and Sunhao Dai and Jingya Dou and Yichen Gong and Yuan Guo and Zhenlin Guo and Feng Li and Qian Li and Jinzhen Lin and Yuqi Zhou and Linchao Zhu and Liang Chen and Zhenyu Guo and Changhua Meng and Weiqiang Wang},
  title        = {UI-Venus-1.5 Technical Report},
  journal      = {CoRR},
  volume       = {abs/2602.09082},
  year         = {2026}
}

@inproceedings{hu2022lora,
  author       = {Edward J. Hu and Yelong Shen and Phillip Wallis and Zeyuan Allen-Zhu and Yuanzhi Li and Shean Wang and Lu Wang and Weizhu Chen},
  title        = {{LoRA}: Low-Rank Adaptation of Large Language Models},
  booktitle    = {International Conference on Learning Representations},
  year         = {2022},
}

@online{openai2026gpt54,
  title        = {Introducing GPT-5.4},
  year         = {Accessed: 2026},
  url          = {https://openai.com/index/introducing-gpt-5-4/}
}

@article{kuhn1955hungarian,
  author       = {Harold W. Kuhn},
  title        = {The {Hungarian} Method for the Assignment Problem},
  journal      = {Naval Research Logistics Quarterly},
  volume       = {2},
  number       = {1-2},
  pages        = {83--97},
  year         = {1955},
}

@article{yang2023som,
  author       = {Jianwei Yang and Hao Zhang and Feng Li and Xueyan Zou and Chunyuan Li and Jianfeng Gao},
  title        = {Set-of-Mark Prompting Unleashes Extraordinary Visual Grounding in {GPT-4V}},
  journal      = {CoRR},
  volume       = {abs/2310.11441},
  year         = {2023}
}

@article{lu2024omniparser,
  author       = {Yadong Lu and Jianwei Yang and Yelong Shen and Ahmed Awadallah},
  title        = {{OmniParser} for Pure Vision Based {GUI} Agent},
  journal      = {CoRR},
  volume       = {abs/2408.00203},
  year         = {2024}
}

@inproceedings{schoop2022tappability,
  author       = {Eldon Schoop and Xin Zhou and Gang Li and Zhourong Chen and Bj{\"o}rn Hartmann and Yang Li},
  title        = {Predicting and Explaining Mobile {UI} Tappability with Vision Modeling and Saliency Analysis},
  booktitle    = {Proceedings of the 2022 CHI Conference on Human Factors in Computing Systems},
  year         = {2022},
}

@inproceedings{wen2024autodroid,
  author       = {Hao Wen and Yuanchun Li and Guohong Liu and Shanhui Zhao and Tao Yu and Toby Jia-Jun Li and Shiqi Jiang and Yunhao Liu and Yaqin Zhang and Yunxin Liu},
  title        = {{AutoDroid}: {LLM}-powered Task Automation in Android},
  booktitle    = {Proceedings of the 30th Annual International Conference on Mobile Computing and Networking},
  year         = {2024},
  pages        = {543--557},
}

@inproceedings{ran2024guardian,
  author       = {Dezhi Ran and Hao Wang and Zihe Song and Mengzhou Wu and Yuan Cao and Ying Zhang and Wei Yang and Tao Xie},
  title        = {{Guardian}: A Runtime Framework for {LLM}-Based {UI} Exploration},
  booktitle    = {Proceedings of the 33rd ACM SIGSOFT International Symposium on Software Testing and Analysis},
  year         = {2024},
  pages        = {958--970},
}

@inproceedings{su2017stoat,
  author       = {Ting Su and Guozhu Meng and Yuting Chen and Ke Wu and Weiming Yang and Yao Yao and Geguang Pu and Yang Liu and Zhendong Su},
  title        = {Guided, Stochastic Model-Based {GUI} Testing of Android Apps},
  booktitle    = {Proceedings of the 2017 11th Joint Meeting on Foundations of Software Engineering},
  year         = {2017},
  pages        = {245--256},
}

@article{wang2025llmdroid,
  author       = {Chenxu Wang and Tianming Liu and Yanjie Zhao and Minghui Yang and Haoyu Wang},
  title        = {{LLMDroid}: Enhancing Automated Mobile App {GUI} Testing Coverage with Large Language Model Guidance},
  journal      = {Proceedings of the ACM on Software Engineering},
  volume       = {2},
  number       = {FSE},
  year         = {2025},
  pages        = {1001--1022},
}

@inproceedings{li2025uipro,
  author       = {Hongxin Li and Jingran Su and Jingfan Chen and Zheng Ju and Yuntao Chen and Qing Li and Zhaoxiang Zhang},
  title        = {{UIPro}: Unleashing Superior Interaction Capability for {GUI} Agents},
  booktitle    = {Proceedings of the IEEE/CVF International Conference on Computer Vision},
  year         = {2025},
  pages        = {1613--1623},
}

@online{qwen2026qwen35_9b,
  title        = {{Qwen3.5-9B}},
  year         = {Accessed: 2026},
  url          = {https://huggingface.co/Qwen/Qwen3.5-9B}
}

@online{qwen2026qwen38_27b,
  title        = {{Qwen3.8-27B}},
  year         = {Accessed: 2026},
  url          = {https://huggingface.co/Qwen/Qwen3.8-27B}
}

@online{qwen2026qwen37plus,
  title        = {{Qwen3.7-Plus}: Multimodal Agent Intelligence},
  year         = {Accessed: 2026},
  url          = {https://qwen.ai/blog?id=qwen3.7-plus}
}

@online{bytedance2026seed21,
  title        = {{Seed2.1}},
  year         = {Accessed: 2026},
  url          = {https://seed.bytedance.com/en/seed2_1}
}

@online{volcengine2026seedevolving,
  title        = {{Seed-Evolving}},
  year         = {Accessed: 2026},
  url          = {https://docs.volcengine.com/docs/ark/latest-model?lang=zh}
}

@online{deepseek2026v4flashvision,
  title        = {{DeepSeek-V4-Flash-Vision-Exp Release}},
  year         = {Accessed: 2026},
  url          = {https://api-docs.deepseek.com/updates/#deepseek-v4-flash-vision-exp-release}
}

@online{openai2026gpt6luna,
  title        = {{GPT-6 Luna}},
  year         = {Accessed: 2026},
  url          = {https://developers.openai.com/api/docs/models/gpt-6-luna}
}

@online{openai2026gpt56luna,
  title        = {{GPT-5.6 Luna}},
  year         = {Accessed: 2026},
  url          = {https://developers.openai.com/api/docs/models/gpt-5.6-luna}
}

\end{document}